\documentclass[11pt]{article}

\usepackage{pdflscape}
\usepackage{amsthm}
\usepackage{floatrow}
\usepackage[margin=1in]{geometry}
\usepackage[comma]{natbib}
\usepackage{anysize}
\usepackage[usenames, dvipsnames]{color}
\usepackage{xcolor,color}
\usepackage{graphicx}
\usepackage{epstopdf}
\usepackage{amsmath,amssymb,amsthm,amsopn,mathrsfs}
\usepackage[linesnumbered,ruled,vlined]{algorithm2e}

\DontPrintSemicolon
\LinesNotNumbered

\newcommand{\nextnr}{\stepcounter{AlgoLine}\ShowLn}

\definecolor{navy}{rgb}{0,0,0.502}

\theoremstyle{definition}
\theoremstyle{plain}\newtheorem{prop}{Proposition}
\theoremstyle{remark}
\newcommand{\tM}{\textrm{M}}
\newcommand{\tO}{\textrm{O}}
\newcommand{\tOO}{\textrm{OO}}
\newcommand{\tMM}{\textrm{MM}}
\newcommand{\tMO}{\textrm{MO}}
\newcommand{\tSM}{\textrm{SM}}
\newcommand{\tSO}{\textrm{SO}}
\newcommand{\bB}{\boldsymbol{B}}
\newcommand{\bS}{\boldsymbol{S}}
\newcommand{\bX}{\boldsymbol{X}}
\newcommand{\bY}{\boldsymbol{Y}}
\newcommand{\bb}{\boldsymbol{b}}
\newcommand{\bi}{\boldsymbol{i}}
\newcommand{\bs}{\boldsymbol{s}}
\newcommand{\bx}{\boldsymbol{x}}
\newcommand{\by}{\boldsymbol{y}}
\newcommand{\bbeta}{\boldsymbol{\beta}}
\newcommand{\real}{\mathbb{R}}
\newcommand{\der}{\mathrm{d}}
\newcommand{\bUpsilon}{\boldsymbol{\Upsilon}}
\newcommand{\bOne}{\bold{1}}
\DeclareMathOperator*{\argmax}{argmax}
\newcommand{\dom}{\mathcal{D}}
\newcommand{\Cov}{\textrm{Cov}}
\newcommand{\Var}{\textrm{Var}}
\newcommand{\Pred}{\textrm{Pred}}
\newcommand{\Model}{\textrm{Model}}

\begin{document}

\title{Logistic regression models for aggregated data}

\author{T. Whitaker\footnote{UNSW Data Science Hub \& School of Mathematics and Statistics, University of New South Wales, Sydney.},\; B. Beranger$^{*}$ and S. A. Sisson$^{*}$}

\date{}

\maketitle 
\theoremstyle{plain}
\newtheorem{mydef}{Proposition}
\theoremstyle{definition}
\newtheorem*{mydef2}{Example}
\theoremstyle{definition}
\newtheorem{mydef3}{Remark}
\theoremstyle{definition}
\newtheorem{mydef4}{Extension}
\theoremstyle{definition}
\newtheorem{mydef5}{Definition}

%
\begin{abstract} 
Logistic regression models are a popular and effective method to predict the probability of categorical response data.
However inference for these models can become computationally prohibitive for large datasets.
Here we adapt ideas from symbolic data analysis to summarise the collection of predictor variables into histogram form, and perform inference on this summary dataset.
We develop ideas based on composite likelihoods to derive an efficient one-versus-rest approximate composite likelihood model for histogram-based random variables, constructed from low-dimensional marginal histograms obtained from the full histogram.
We demonstrate that this procedure can achieve comparable classification rates compared to the standard full data multinomial analysis and against state-of-the-art subsampling algorithms for logistic regression, but at a substantially lower computational cost. 
Performance is explored through simulated examples, and analyses of large supersymmetry and satellite crop classification datasets.
\\

\\
 
%
%
%
\end{abstract}
%

\section{Introduction}

There are many well developed statistical methods for classification, such as logistic regression, discriminant analysis and clustering, which predict a categorical variable that can take one of $K$ distinct values 
given an input vector of predictor variables
\citep[e.g.][]{hastie, pampel2000}. While these methods are effective for the analysis of standard data, 
when the data take non-standard forms, such as random interval- or random histogram-based predictors,
existing methods are either underdeveloped or do not exist.
Interval, histogram and other-distribution based data summaries can arise through measurement error, data quantisation, expert elicitation and, motivating this work, the desire to summarise large and complex datasets in an appropriate way so that they can be  analysed more efficiently than the full dataset \citep[e.g.][]{zhang+bs20}. 
The field of Symbolic Data Analysis (SDA) was developed to analyse such distributional data
\citep{diday1989, billard+d03, billard2011,beranger+ls18}. However, with a few exceptions (discussed below), the parameters of existing SDA-based methods 
undesirably lose their interpretation as parameters of models of the underlying (standard) micro-data, which means that the resulting inferences are not directly comparable to the equivalent standard full-data analysis \citep{zhang+bs20,beranger+ls18}.

Logistic regression \citep[e.g.][]{cox,logbook}  is one method of performing regression for categorical response data, and has been utilised extensively in many fields, including finance \citep{hauser2011, Kim2010}, epidemiology \citep{Merlo2006}, medicine \citep{Min2013, logbook}, diagnostics \citep{Knottnerus1992} and modelling  income \citep{Pavlopoulos2010}. 
Such models are typically fitted by numerical maximum likelihood estimation as there is no closed form maximum likelihood estimator (MLE),  conditional on certain conditions on the response and predictors such that the MLE exists \citep{Adelin1984}.
As a result, the computational overheads for determining the MLE of logistic regression models can be high for very large datasets

With this motivation, \cite{wang2018} developed a data-subsampling scheme for binary logistic regression
(for $K=2$ classes) whereby each observation is assigned a weight according to a function that minimises the asymptotic mean squared error (MSE) of the MLE. A subsample is then taken from the full dataset using these weights, with parameter estimates then obtained by maximising the likelihood for the subsample. To calculate the observation weights, an initial parameter estimate is obtained using  $r_0$ uniformly sampled observations. \cite{wang2018} demonstrated good estimator performance from the `optimal' subsample using $r_0=1\,000$.

In contrast, \cite{deSouza2} presented SDA-motivated versions of logistic regression for interval-valued data (where the intervals are constructed from the minimum and maximum observed values of each univariate predictor in the micro-data), whereby the regressions are constructed using the interval centres or endpoints as covariates.
\cite{deSouza1} developed a similar approach for multi-class classification with logistic regression using interval endpoints.
While these SDA methods are computationally simple, because they are only based on random intervals (that is, two quantiles of the data) much of the distributional information in the predictor values is lost. In addition the implementations treat each interval equally, which can cause problems if there are unbalanced numbers of micro-data in each category. Finally, as discussed above, the fitted models cannot be compared to the equivalent models fitted to the full (non-summarised) dataset.

In other work, \cite{steel1997} investigated the effect of data aggregation on logistic regression when both the predictor and response variables are observed as the total sum of each variable for each group. \cite{genagg} proposed an iterative algorithm for estimating Generalised Linear Models (GLMs) 
when the response variables are aggregated into histogram-valued random variables. 
The resulting training error was numerically shown to approach that of the full (non-aggregated) analysis as the number of histogram bins became large.
\cite{Armstrong1985} considered the case where a single covariate is measured with error, and the distribution of the coarsened (binned) value given the true latent observation is known. Estimates for GLM coefficients are then obtained through the utilisation of this known density.
\cite{johnson2006} assumed Gaussian distributions for coarsened 
covariates and included their likelihood within the GLM framework. 
\cite{lipsitz2004} proposed a method for implementing a GLM when one of the covariates is coarsened for only a subset of observations, whereby the resulting likelihood is the integral over the likelihood of all possible values that this variable could have taken, weighted by a density dependent on the uncoarsened, fully observed variables. 
\cite{johnson2014} proposed a Bayesian approach for coarsened covariates in GLMs whereby a 
distribution is assumed for the coarsened covariates, given the uncoarsened data. 

The above models work well if the distribution of the latent data given the observed data is well-specified, but run into problems if the distributional assumptions are  
violated. 
Further, if many covariates are provided in distributional form, these approaches require large computational overheads due to the curse of dimensionality. To the best of our knowledge, little progress has been made in developing logistic regression models for fully histogram-valued predictor variables, which provide far more insight into the shape of the underlying covariate data than the interval case. Here, our primary motivation is in histograms constructed from very large datasets by the analyst in order to facilitate increased computational speed of an analysis. 
In this setting, data could arrive in the form of univariate histograms for each predictor for each of the $K$ categories, instead of large $(N\times (K+1))$-dimensional tables (where $N$, the number of observed predictor and class label pairs, is very large), allowing savings in data transmission, storage and analysis.

In this article we develop methods for implementing logistic regression models with $K$ response categories,
where the covariate data for each response category is in the form of marginal or multivariate histograms (or random rectangles, as a special case of random histograms with a single bin).
The basic component of our approach adopts the likelihood-based SDA construction of \cite{beranger+ls18}  \citep[see also][]{zhang+bs20,whitaker2020}, which unlike other SDA-based methods, directly fits models for the underlying micro-data given the distributional-based data summaries.  For this approach, suppose we let the standard classical likelihood be denoted by $L(\bx, y;\theta)\propto\prod_{n=1}^Ng_{X,Y}(y_n|x_n,\theta)$ for $N$ i.i.d.~pairs $(x_n,y_n)$, where $g_{X,Y}(y_n|x_n,\theta)$ is the assumed distribution of the categorical response $Y$, given a set of covariates $X$ and a parameter $\theta$. Suppose that a histogram of the covariate values $(x_1,\ldots,x_n)$ is observed for each response category, rather than the observed values (the $x_k$) themselves, so that it is only known how many covariate vectors, $s_b$, are contained within bin volume $\Upsilon_b$, for each bin $b$ in the histogram.
The main idea is to obtain the aggregated likelihood by 
averaging the unknown location of the covariate vector in the classical likelihood uniformly over the histogram bin in which it resides.
As a result, the likelihood contribution of a given bin $b$ is proportional to $\left (\int_{\Upsilon_b}g_{X,Y}(y_n|x_n,\theta)dx_n\right )^{s_b}.$

We propose a mixed model in which the underlying data is analysed using a mixture of standard and SDA likelihoods: histogram bins with low counts are discarded and the residing micro-data are used directly to contribute standard likelihood terms, while bins with high counts contribute via the SDA likelihood framework. 
However, due to the computational difficulties associated with the potentially high-dimensional integrations required within the SDA likelihood (i.e.~$\left (\int_{\Upsilon_b}g_{X,Y}(y_n|x_n,\theta)dx_n\right )^{s_b}$), this mixed model is unsuitable for moderate numbers of predictor variables. 
For this reason, along with the benefits accompanying the potential additional savings in data storage and computational overheads, we develop models based on lower-dimensional marginal histograms. 

We introduce an approximate marginal composite likelihood approach to obtain estimates for the parameter vector of the complete (micro-data) model, using lower-dimensional  marginal histograms of the observed covariate data. Univariate and bivariate \textcolor{red}{covariate} histograms are considered for these models, with the predictions performed on test datasets shown to be comparable with the full standard likelihood models, but at a much lower computational cost. 
The resulting parameter estimates are directly comparable to the parameter estimates of the standard full-data analysis.

This article is structured as follows. In Section \ref{sec:log_reg} we recap two types of established logistic regression models for regular data. In Section \ref{sec:sym} we briefly outline the general SDA-based likelihood construction of \cite{beranger+ls18}, and introduce the mixed standard/SDA likelihood model. We also develop the approximate composite likelihood approach based on lower-dimensional marginal histograms.
In Section \ref{sec:sim} we perform various simulation studies to demonstrate the effectiveness of each model. We show that each model is able to produce comparable prediction accuracy compared to the full model at a lower computational cost for a certain sample size, $N$. We also show that our method performs comparably with the recently developed optimal subsampling method for binary logistic regression by \cite{wang2018}, but at a cheaper computational cost. In Section \ref{realdata} we analyse satellite crop-prediction data from Queensland, Australia, and simulated particle collision dataset from the Machine Learning Repository \citep{dua2019}, and show that our approach is highly competitive with  much more computationally expensive standard statistical methods. We conclude with a discussion.


\section{Logistic regression methods for classification}
\label{sec:log_reg}

There are a variety of methods for classifying an instance into one of $K \geq 2$ classes. 
Let $Y$ denote a discrete random variable taking a value in $\Omega= \{1,\ldots,K\}$ and $X \in \real^D$ an associated vector of explanatory variables. Using the information contained in the covariates $X$, the aim is to estimate the probability of the outcome of $Y=k$ for $k \in \Omega$. 
Logistic regression is a widely used statistical modelling technique for binary ($K=2$) dependent variables. Multinomial logistic regression is a generalisation of logistic regression to problems with possible outcomes taking values in $\Omega$ ($K\geq 2$).
An alternative problem representation recasts multinomial classification as multiple binary classification problems. One-vs-Rest (OvR) logistic regression implements a separate binary logistic regression for each class $k \in \Omega$, assuming that each classification model is independent. To establish notation, we briefly describe both multinomial and OvR logistic regression methods below.

\subsection{Multinomial logistic regression}
\label{ssec:multinomial}

For each outcome $k \in \Omega\backslash\{K\}$, the multinomial logistic regression model defines the log pairwise odds ratios through a linear model
\begin{equation}
\log \left (\frac{P_{\tM}(Y=k| X)}{P_{\tM}(Y=K| X)}\right ) = \beta_{k0} + \beta_k^\top  X,
\label{eq:logodds}
\end{equation}
where $P_{\tM}(Y=k | X)$ denotes the probability that $Y=k$, under the multinomial model (M),  when $X$ is observed, $\beta_{k0} \in \real$ is an intercept and $\beta_k = (\beta_{k1}, \ldots, \beta_{kD})^\top\in \real^D$ represents the vector of regression coefficients associated with the $D$ explanatory variables and the outcome $k$.
The outcome $K$ is referred to as the pivot or reference category, and its corresponding parameter $(\beta_{K0},\beta_K^\top)^\top$ is the zero vector. Thus \eqref{eq:logodds} can be rearranged as
$$
P_{\tM} (Y=k | X) = \frac{e^{\beta_{k0} + \beta_k^\top  X}}{1+\sum_{j \in \Omega\backslash\{K\}} e^{\beta_{j0} + \beta_j^\top  X}}
$$
for all $k \in \Omega$. This implies that 
the odds of preferring one class over another do not depend on the presence or absence of other ``irrelevant" alternatives.

Suppose that $\bX=( X_1,..., X_N)$ is a vector of $D$-dimensional random vectors composed of $N$ i.i.d.~replicates of $X$ and $\bY= (Y_1,...,Y_N)^\top$ is a vector of $N$ i.i.d.~replicates of $Y$,  with respective realisations $\bx \in \real^{D\times N}$ and $\by \in \Omega^N$. The likelihood function under the multinomial model is given by
\begin{equation}
L_{\tM} (\bx, \by; \bbeta) = \prod_{n=1}^N \prod_{k \in \Omega} P_{\tM} (Y = k | X=x_n)^{\boldsymbol{1} \{y_n=k\}},
\label{eq:MLR}
\end{equation} 
where $\boldsymbol{1}\{\cdot\}$ represents the indicator function, and $\bbeta = (\check{\beta}_1, \ldots, \check{\beta}_K) \in \real^{(D+1)\times K}$ with $\check{\beta}_k = (\beta_{k0}, \beta_k^\top)^\top \in \real^{D+1}$. 
We denote the maximun likelihood estimator for the multinomial logistic regression model as $\hat \bbeta^\tM = \argmax_{\bbeta} \log L_{\tM}(\bx, \by;\bbeta)$.
Existence of the MLE can be examined through the concept of data separation.
\begin{mydef5}\label{def1}{\bf (Multinomial model separation)}
There is complete separation of the data if for all $k\in\Omega$, a $\bb=(b_1,\ldots,b_K)$, $b_k\in\mathbb{R}^D$,  exists such that 
\begin{align*}
(b_k-b_j)^\top x_n& > 0 \mbox{ for all } n \mbox{ such that } y_n=k, j\neq k\\
(b_k-b_j)^\top x_n& < 0 \mbox{ for all } n \mbox{ such that } y_n\neq k, j\neq k.
\end{align*}
There is quasi-complete separation of the data if for all $k\in\Omega$, a $\bb=(b_1,\ldots,b_K)$, $b_k\in\mathbb{R}^D$,  exists such that 
\begin{align*}
(b_k-b_j)^\top x_n& \geq 0 \mbox{ for all } n \mbox{ such that } y_n=k, j\neq k\\
(b_k-b_j)^\top x_n& \leq 0 \mbox{ for all } n \mbox{ such that } y_n\neq k, j\neq k,
\end{align*}
with equality for at least one observation $x_n$ in each class $k$. 
\end{mydef5}
\cite{Adelin1984} proved that if there is neither complete nor quasi-complete separation in the data, then the MLE $\hat{\bbeta}^{\tM}$ exists for 
all classes $k\in\Omega$.

\subsection{One-vs-Rest logistic regression}
\label{ssec:OvR}

For each outcome $k \in \Omega$, the One-vs-Rest logistic regression model defines the log odds ratio through a linear model
$$
\log \left (\frac{P_{\tO}(Y=k| X)}{P_{\tO}(Y\neq k| X)}\right ) = \beta_{k0} + \beta_k^\top  X, 
$$
where $P_{\tO}(Y=k| X)$ denotes the probability that $Y=k$, under the  OvR regression model ($\tO$), when X is observed and where $\beta_{k0}$ and $\beta_k $ are as defined previously.
This ratio can be rearranged as
$$
P_{\tO}(Y=k | X) = \frac{e^{\beta_{k0} + \beta_k^\top  X}}{1+e^{\beta_{k0} + \beta_k^\top  X}} $$
for all $k \in \Omega$.
Note that here $\beta_K$ is different from the zero vector as each individual binary model has an implied reference category and $\sum_{k \in \Omega} P_\tO (Y=k| X)\neq 1$. The likelihood function can be written as
\begin{align}
\label{eq:OVR}
L_\tO (\bx, \by; \bbeta) 
&= \prod_{n=1}^N \left( P_\tO (Y=y_n | X=x_n)
\prod_{k\in \Omega \backslash\{y_n\}} P_\tO (Y \neq k | X=x_n)
\right),
\end{align}  
which is expressed as the product of 
$K$ binary logistic regressions for each observation $(x_n, y_n)$. The parameters of the multivariate and OvR regression models are not directly comparable, but the performance of each model can be assessed by evaluating their prediction accuracy on a training dataset \citep[e.g.][]{eichelberger2013}. 

As before,  the MLE under the OvR model,  $\hat \bbeta^\tO = \argmax_{\bbeta} \log L_{\tO}(\bx, \textcolor{red}{\by};\bbeta)$, exists for all classes $k\in\Omega$ if there is neither complete nor quasi-complete separation of the data for each $k\in\Omega$, but under slightly modified definitions of separation compared to the multinomial model  \citep{Adelin1984}.
\begin{mydef5}\label{def2}{\bf (OvR model separation)}
There is complete separation of the data  for the $k^{th}$ class if a $b_k\in\mathbb{R}^D$ exists such that 
\begin{align*}
b_k^\top x_n& > 0 \mbox{ for all } n \mbox{ such that } y_n=k\\
b_k^\top x_n& < 0 \mbox{ for all } n \mbox{ such that } y_n\neq k.
\end{align*}
There is quasi-complete separation of the data for the $k^{th}$ class if a $b_k\in\mathbb{R}^D$ exists such that 
\begin{align*}
b_k^\top x_n& \geq 0 \mbox{ for all } n \mbox{ such that } y_n=k\\
b_k^\top x_n& \leq 0 \mbox{ for all } n \mbox{ such that } y_n\neq k,
\end{align*}
with equality for at least one observation $x_n$ in the $k^{th}$ class, and at least one observation $x_{n'}$, $n'\neq n$, not in the $k^{th}$ class. 
\end{mydef5}

\section{Classification for aggregated data }  
\label{sec:sym}

For each class $k \in \Omega$, define $\bX^{(k)} = \left( X_n | Y_n =k, n=1,\ldots,N \right) \in \real^{D\times N_k}$, where $N_k = \sum_{n=1}^N \boldsymbol{1} \{ Y_n = k \}$, as the matrix of $N_k$ covariate vectors associated with  outcome $k$. 
In this way $\bX^{(1)},\ldots,\bX^{(K)}$ partition the full covariate set $\bX$.
When $N$ is very large and $|\Omega| \ll N$ then it can be expected than one or more of the $N_k$ will also be large. In this context, directly optimising likelihood functions for logistic regression models could be computationally prohibitive. As an alternative,
 it might be appealing to aggregate the information contained in $\bX^{(k)}$ into distributional form (such as a histogram) and to implement a classification algorithm using these summaries only, which could be much more computationally efficient. The concept of performing statistical or 
 inferential analyses on such distributional ``datapoints'' originated from \cite{diday1989}, and has become known as symbolic data analysis, where the `symbol' corresponds to the distributional summary \citep[see also][]{billard+d03, billard+d06, bock2000}. 

A symbolic random variable $S_k \in \dom_{S_k}$ can be viewed as the result of applying an aggregation function $\pi (\cdot)$ to $\bX^{(k)} \in \dom_{\bX^{(k)}}$ (where $\dom_{\bX^{(k)}}= \real^{D\times N_k}$), i.e.~$S_k = \pi( \bX^{(k)}): \dom_{\bX^{(k)}} \rightarrow \dom_{S_k}$ so that $\bx^{(k)} \mapsto s_k$. In the present context $s_k$ corresponds to a vector of counts of the number of covariate vectors in $\bx^{(k)}$ that reside in each histogram bin (see Section \ref{ssec:hist_LR} below for more explicit detail).
Various likelihood-based techniques for fitting statistical models given the information in distributional summaries
have been developed \citep{lerademacher+b10,brito+d12,lin+cs17,beranger+ls18,zhang+bs20}. 
Here we follow the construction of \citet{beranger+ls18} and \cite{zhang+bs20} who fully model the construction of the symbols from the generating process of the standard random variables $\bX^{(k)}$. 
Specifically, the likelihood of observing $s_k$ is
\begin{equation}
L(s_k; \theta, \vartheta) \propto 
\int_{\dom_{\bX^{(k)}}} f_{S_k | \bX^{(k)} = \bx^{(k)}} \left(s_k | \bx^{(k)}, \vartheta \right) 
g_{\bX^{(k)}} \left(\bx^{(k)}; \theta\right) \der \bx^{(k)},
\label{eq:symboliclik}
\end{equation}
where $f_{S_k | \bX^{(k)}}(\,\cdot\,;\vartheta)$ is the conditional density of $S_k$ given $\bX^{(k)}$ relating to the aggregation of $\bx^{(k)} \mapsto s_k$, $g_{\bX^{(k)}} (\bx^{(k)}; \theta)$ is the standard likelihood function of the model at the data level with parameter of interest $\theta$, and $\bx^{(k)} = \left(x_1^{(k)}, \ldots, x_{N_k}^{(k)}\right)$, with $x_n^{(k)} \in \real^D$ denoting the covariate vector of the $n$-th observation with outcome $k$.
The likelihood \eqref{eq:symboliclik} is a general expression, for which the density $f_{S_k | \bX^{(k)}}(\,\cdot\,;\vartheta)$ takes different forms depending on the type of distributional summary considered \cite[see][for several examples using random intervals/rectangles and random histograms]{beranger+ls18}.
Consistency results for the above likelihood function have been provided by \cite{rahman+brs20}, in a particular application setting.
In the following we are interested in aggregating the covariates $\bX^{(k)}$ that have the same outcome $k$ into histograms (with fixed or random bins), $S_k$, and to fit logistic regression type models ($g_{\bX^{(k)}} \left(\bx^{(k)}; \theta\right)$).

\subsection{Logistic regressions using histogram-valued data}
\label{ssec:hist_LR}

For each class $k \in \Omega$, suppose that the $d$-th margin of $\real^D$ is partitioned into $B_k^d$ bins, so that $B_k^1 \times \ldots \times B_k^D$ bins are created in $\real^D$ through the $D$-dimensional intersections of each marginal bin. Index each bin by $\bb_k = (b_{1_k}, \ldots, b_{D_k})$, $b_{d_k} = 1, \ldots, B_k^d$, as the $D$-dimensional vector of co-ordinates of each bin in the histogram. The bin $\bb_k$ is constructed over the space $\bUpsilon_{\bb_k} =  \Upsilon_{\bb_k}^{1}\times \cdots \times \Upsilon_{\bb_k}^D\subset\mathbb{R}^D$, where $\Upsilon_{\bb_k}^d = (y_{b_{d}-1}^d,y_{b_d}^d]\subset\real$ is a univariate bin in the $d$-th margin, and where, for each margin $d$,
$-\infty<y^d_0<y^d_1<\ldots<y^d_{B^k}<\infty$ are fixed points that define the change from one bin to the next. The index $k$ has been omitted in the above bin delimitations in order not obscure notation any further, but it needs to be kept in mind that these are specific to the outcome $k \in \Omega$.

Let $S_k$ represent a $D$-dimensional histogram constructed from $\bX^{(k)}$ through the aggregation function $\pi$ where
\begin{align}
\label{eq:hist_const}
S_k = \pi(\bX^{(k)}): \real^{N_k \times D} & \rightarrow \{0, \ldots ,N_k \}^{B^1_k \times \ldots \times B_{k}^D}\\
\bx^{(k)} &\mapsto s_k = \left (s_{\bOne_k} = \sum_{n=1}^{N_k} \boldsymbol{1}\{  x_n^{(k)} \in \bUpsilon_{\bOne_k} \}, \ldots, s_{\bB_k}=\sum_{n=1}^{N_k} \boldsymbol{1}\{ x_n^{(k)} \in \bUpsilon_{\bB_k}\} \right ) \nonumber.
\end{align}
The quantity $S_{\bb_k}$ denotes the random number of observed data points $X_1^{(k)}, \ldots, X_{N_k}^{(k)}$ that fall in the bin indexed by $\bb_k$. Consequently, the histogram-valued random variable $S_k = (S_{\bOne_k}, \ldots, S_{\bB_k})$ represents the full $(B_k^1 \times \ldots \times B_k^D)$-dimensional vector of counts from the first bin $\bOne_k = (1,\ldots, 1)$ to the last bin $\bB_k = (B_k^1, \ldots, B_k^D)$. 
In this manner, we can construct the collection of histograms $\bS=(S_1,\ldots,S_K)$, with one $S_k$ for each outcome index $k\in\Omega$, that summarise the information contained in $\bX=(\bX^{(1)},\ldots,\bX^{(K)})$.

\begin{prop}
\label{prop1}
Suppose that $\bX^{(k)}$, the covariates associated with each outcome $k \in \Omega$, are aggregated via \eqref{eq:hist_const}, and let $\bS = (S_1, \ldots, S_K)$ denote the resulting collection of histograms. 
For this summarised data $\bS$, using \eqref{eq:symboliclik}, the likelihood functions for the multinomial \eqref{eq:MLR} and OvR \eqref{eq:OVR} logistic regression models become 
\begin{align}
\label{eq:SM}
L_{\tSM} (\bs; \bbeta) & \propto \prod_{k\in\Omega} 
\prod_{\bb_k=\bOne_k}^{\bB_k} \left(\int_{\bUpsilon_{\bb_k}} P_{\tM} (Y=k | X=x) \der x \right )^{s_{\bb_k}} \\
\label{eq:SO}
L_{\tSO}( \bs;\bbeta) & \propto \prod_{k \in \Omega} \prod_{\bb_k= \bOne_k}^{\bB_k} 
\left(
\int_{\bold \Upsilon_{\bb_k}}P_{\tO}(Y=k| X=x) \der x
\prod_{k' \in \Omega\backslash \{k\} } 
\int_{\bold \Upsilon_{\bb_k}}P_{\tO}(Y\neq k'| X=x) \der x
\right)^{s_{\bb_k} },
\end{align}
where $\bs$ is the observed value of $\bS$, and $s_{\bb_k}$ denotes the number of observations in bin $\bb_k$ in histogram $k$.  For a derivation 
see Appendix \ref{proofprop1}.
\end{prop} 

We refer to these models as the symbolic multinomial (SM) and symbolic One-vs-Rest (SOvR) logistic models.
In effect, the uncertainty of the location of each predictor $X$ is averaged uniformly over its location in the histogram bin in which it resides.
Note that the likelihood functions \eqref{eq:SM} and \eqref{eq:SO} only implicitly depend on the vector of outcomes \textcolor{red}{$\by$}
since for each possible outcome $k \in \Omega$ the covariates $\bX^{(k)}$ are summarised in a histogram $S_k$, and so the product of the $N$ $(Y_n,X_n)$ observations in the standard likelihoods (\eqref{eq:MLR} and \eqref{eq:OVR}) is replaced by a product over the $K$ outcomes. 
Further, the  parameter $\vartheta$ in \eqref{eq:symboliclik} denotes quantities relevant to constructing the symbol (e.g.~the number of bins and their locations), and so is fixed in this setting and is therefore omitted in the notation. 

Following similar arguments to \citet{heit1989, beranger+ls18}, the symbolic likelihoods $L_{\tSM}(\bs;\bbeta)$ and $L_{\tSO}(\bs;\bbeta)$ can each be shown to approach their classical equivalent,  $L_{\tM}(\bx, \by;\bbeta)$ and $L_{\tO}(\bx,\by;\bbeta)$, as the number of bins in each histogram approaches infinity and the volume of each bin approaches zero. In this scenario, in the limit each bin will either be empty ($s_{\bb_k}=0$) and so the bin will not contribute a likelihood term, or will contain exactly one point ($s_{\bb_k}=1$) observed at each value of $x^{(k)}_n$, which then recovers the classical likelihood term for $x^{(k)}_n$.
In this manner, the histogram-based likelihoods can be viewed as approximations to the standard likelihood functions for each model.

To establish conditions for the existence of the respective MLEs, $\hat{\bbeta}^{\tSM}=\arg\max_{\bbeta} \log L_{\tSM} (\bs; \bbeta)$ and $\hat{\bbeta}^{\tSO}=\arg\max_{\bbeta} \log L_{\tSO} (\bs; \bbeta)$, we need to consider modified definitions of complete and quasi-complete separation of the data, in analogy with Definitions \ref{def1} and \ref{def2}, to account for the fact that the location of each covariate vector $x_n$ is only known up to the histogram bin in which it resides.

\begin{mydef5}\label{def3}{\bf (Histogram-based multinomial model separation)}
There is complete separation of the 
set of histograms $\bs$ if for all $k\in\Omega$, a $\bb=(b_1,\ldots,b_K)$, $b_k\in\mathbb R^D$, exists such that 
\begin{align*}
(b_k-b_j)^\top x& > 0 \mbox{ for all } x\in \bUpsilon_{\bb_k} \mbox { such that } s_{\bb_k}>0, j\neq k\\
(b_k-b_j)^\top x& <0 \mbox{ for all } x\in \bUpsilon_{\bb_{k'}}\mbox { such that } s_{\bb_{k'}}>0,j\neq k\mbox{ and }k'\neq k.
\end{align*}
There is quasi-complete separation of the 
set of histograms $\bs$ if for all $k\in\Omega$,  a $\bb=(b_1,\ldots,b_K)$, $b_k\in\mathbb R^D$, exists such that 
\begin{align*}
(b_k-b_j)^\top x& \geq 0 \mbox{ for all } x\in \bUpsilon_{\bb_k} \mbox { such that } s_{\bb_k}>0, j\neq k\\
(b_k-b_j)^\top x& \leq 0 \mbox{ for all } x\in \bUpsilon_{\bb_{k'}}\mbox { such that } s_{\bb_{k'}}>0,j\neq k\mbox{ and }k'\neq k.
\end{align*}
with equality for at least one point in the non-empty histogram bins for the $k$-th class.
\end{mydef5}

\begin{mydef5}\label{def4}{\bf (Histogram-based OvR model separation)}
There is complete separation of the 
histogram for the $k^{th}$ class, $s_k$, if  a vector $b_k$ exists such that 
\begin{align*}
b_k^\top x& >0 \mbox{ for all } x\in \bUpsilon_{\bb_k} \mbox { such that } s_{\bb_k}>0\\
b_k^\top x& <0 \mbox{ for all } x\in \bUpsilon_{\bb_{k'}}\mbox { such that } s_{\bb_{k'}}>0, k\neq k'.
\end{align*}
There is quasi-complete separation of the 
histogram for the $k^{th}$ class if there exists a vector $b_k$ such that 
\begin{align*}
b_k^\top x& \geq 0 \mbox{ for all } x\in \bUpsilon_{\bb_k} \mbox { such that } s_{\bb_k}>0\\
b_k^\top x& \leq 0 \mbox{ for all } x\in \bUpsilon_{\bb_{k'}}\mbox { such that } s_{\bb_{k'}}>0,k\neq k'.
\end{align*}
with equality for at least one point in the non-empty histogram bins for the $k$-th class, and equality for at least one point in any non-empty histogram bin not in the $k$-th class.
\end{mydef5}

 \begin{prop}\label{prop:NewProp}
 If the set of histograms $\bs=(s_1,\ldots,s_K)$ does not exhibit complete or quasi-complete separation as described under Definition \ref{def3}, then $L_{\tSM}( \bs;\bbeta) $ has a unique global maximum. If the set of histograms does not exhibit complete or quasi-complete separation for any class $k\in\Omega$ as described under Definition \ref{def4}, then $L_{\tSO}( \bs;\bbeta) $ has a unique global maximum. 
For a proof see Appendix \ref{SeparationProp}.
\end{prop}

From Definitions \ref{def1}--\ref{def4} it can be seen that if there is separation of the histograms then there also has to be separation of the underlying data. In this sense, the definitions of complete and quasi-complete separation for random histograms (Definitions \ref{def3} and \ref{def4}) are stronger conditions than those for standard random vectors (Definitions \ref{def1} and \ref{def2}). As a result, from Proposition \ref{prop:NewProp} this means that if no histogram-based MLE ($\hat{\bbeta}^{\tSM}$ or $\hat{\bbeta}^{\tSO}$) exists, then the equivalent standard logistic model MLE ($\hat{\bbeta}^{\tM}$ or $\hat{\bbeta}^{\tO}$) also does not exist. That is, the standard MLE can't exist without the histogram-based MLE also existing.
Conversely, however, it is possible to have separation of the underlying data but no separation in the derived histograms as a result of covariate vector's locations being `blurred' in the histogram bins. Accordingly, it is possible that the histogram-based MLE exists without the standard logistic model MLE existing.  
(In this particular setting we therefore have the interesting case of the existence of an MLE for a given histogram converging to the non-existence of the MLE in the limit as the number of histogram bins become large while the volume of each bin approaches zero.). 

It is also possible for the histogram-based  MLEs to exist for one histogram derived from an underlying dataset, but not exist for a different histogram (e.g.~with different numbers of and/or locations of bins) derived from the same dataset. For example, a shift in the location of the bins can lead to low-count bins becoming empty (or empty bins becoming non-empty), meaning that a set of histograms previously exhibiting separation, could instead exhibit non-separation (and vice versa). Practically, as a result, if the underlying classical data is available,  the practitioner should consider verifying the separation conditions on the underlying dataset (Definitions \ref{def1} and \ref{def2}) prior to aggregation, to ensure the appropriateness of the logistic model.

Assuming that the MLEs exist, the benefits of using histograms as data summaries for logistic regression modelling are obvious in the presence of large amounts of data. The effective number of likelihood terms in $L_{\tSM} (\bs; \bbeta)$ and $L_{\tSO} (\bs; \bbeta)$ is the number of histogram bins multiplied by the number of classes. For very large datasets this can be much smaller than the $N$ terms in $L_{\tM} (\bx, \by; \bbeta)$ and $L_{\tSO} (\bx, \by; \bbeta)$, and so computing MLEs given the histogram summaries can be much more efficient. The trade off is the loss of some accuracy due to the loss of information in the binned data. 

 Despite its computational advantages, the above construction has some limitations.
 We now discuss these and propose some statistical and computational improvements.

\subsection{Using both classical data and histograms}
\label{sec:stat_improv}

When constructing histograms, for example using the method described in Section \ref{ssec:hist_LR}, the number of observations $s_{\bb_k}$ within each bin $\bb_k$ will typically vary widely over bins, from very low to very high counts. Where bins have high counts, large computational efficiencies are obtained in the evaluation of $L_{SM}(\bs;\bbeta)$ and $L_{SO}(\bs;\bbeta)$ over the standard logistic regression likelihood functions.
However, when a bin has low numbers of underlying data points, it may be that the computational cost in evaluating the bin-specific integrals in \eqref{eq:SM} and \eqref{eq:SO} (such as e.g.~$\int_{\bUpsilon_{\bb_k}} P_{\tM} (Y=k | X=x) \der x$) is just as high or higher than evaluating the standard likelihood contributions (e.g.~$P_{\tM} (Y=k | X=x_n)$) for each of the underlying datapoints in that bin. Taken together with the loss of information in moving from the underlying data to a count of datapoints in a bin, in this case it is obviously worse statistically (and perhaps also computationally) to work with the histogram bin rather than the original data in that bin. Creating bins with low data counts becomes more likely as the dimension $D$ of the predictors increases.

To avoid this situation, we introduce a lower bound, $\tau_k\in\{1,\ldots, N_k\}$, on the number of underlying datapoints in a bin region $\bUpsilon_{\bb_k}$ that is required before these data can be summarised into a histogram bin for their contribution to the likelihood function. When the number of underlying datapoints is lower than $\tau_k$, the original data $x_n$ are retained, and contribute to the likelihood in the standard way.  Selection of the threshold $\tau_k$ can be 
based on 
the time needed to evaluate the likelihood for the binned data (of either the exact value or a numerical approximation of the integral in \eqref{eq:SM} or \eqref{eq:SO}) compared to the standard likelihood function, so that binned data are used when it is more computationally efficient to do so.

Under the assumption that the underlying data are available (which may not always be the case), and that the evaluation of the classical likelihood is computationally infeasible given the data size, we therefore propose to use the modified aggregation function 
\begin{align*}
S_k = \tilde{\pi}(\bX^{(k)}):  \real^{N_k\times D} &\rightarrow \{\tau_k, \ldots,N_k \}^u \times \real^{v \times D} \\
\bx^{(k)} &\mapsto 
\left(
\left\{
\begin{array}{lc}
s_{\bb_k} = \sum_{n=1}^{N_k} \boldsymbol{1}\{  x_n^{(k)} \in \bUpsilon_{\bb_k} \} & \textrm{if } s_{\bb_k} \ge \tau_k \\
\bx^{(k)}_{\bb_k} = 
\{  x_n^{(k)}: x_n^{(k)} \in \bUpsilon_{\bb_k} \} & \textrm{otherwise}
\end{array},
\right.
\bb_k = \bOne_k, \ldots, \bB_k
\right),
\end{align*}
where $\tau_k \in \{ 1, \ldots, N_k\}$, $u \in  [0, \ldots, B_k^1 \times \ldots \times B_k^D ]$ is the number of histogram bins containing at least $\tau_k$ observations, and $v= N_k - \sum s_{\bb_k}$ is the number of retained classical datapoints in bins containing less than $\tau_k$ observations. 
That is, the resulting $S_k$ is a mixture of those histogram bins that contain at least $\tau_k$ observations, combined with any remaining predictor vectors $X_n^{(k)}$ that would otherwise be put into bins with less than $\tau_k$ observations.

In the context of logistic regression modelling, this mixture histogram construction produces likelihood functions that are a mixture of the standard and histogram-based likelihood functions 
given in Sections~\ref{ssec:multinomial}, \ref{ssec:OvR} and \ref{ssec:hist_LR}. For example, we can construct the likelihood function for a mixture of histogram and classical data under the multinomial logistic regression model as
\begin{equation}
\label{eqn:MMlik}
L_{\tMM}(\bs; \bbeta) \propto \prod_{k \in \Omega} \prod_{\bb_k= \bOne_k}^{\bB_k}
\left(\int_{\bUpsilon_{\bb_k}} P_{\tM} (Y=k | X=x) \der x \right )^{s_{\bb_k} \boldsymbol{1}\{ s_{\bb_k}\geq\tau_k \}} 
\left( \prod_{x \in \bx^{(k)}_{\bb_k} }  P_{\tM} (Y=k | X=x) \right)^{\boldsymbol{1}\{  s_{\bb_k}<\tau_k \}},
\end{equation}
where $\tMM$ denotes the multinomial mixture. 
A similar mixture-likelihood, $L_{\tMO}( \bs;\bbeta)$, can be constructed for the OvR logistic regression model.

Because $L_{\tMM}(\bs;\bbeta)$ can be considered as a special case of $L_{\tSM}(\bs;\bbeta)$ (and $L_{\tMO}(\bs;\bbeta)$  a special case of $L_{\tSO}(\bs;\bbeta)$) in which the retained classical data vectors $x_n$ can be thought of as residing in zero-volume bins, one for each retained vector, it is immediate that $L_{\tMM}(\bs;\bbeta)\rightarrow L_{M}(\bx,\by;\bbeta)$ also approaches that classical data likelihood function as the number of bins becomes large and the volume of each bin approaches zero (similarly $L_{\tMO}(\bs;\bbeta)\rightarrow L_{O}(\bx,\by;\bbeta)$).

Similarly, by considering the obvious definition of complete and quasi-complete separation for the mixture of histogram and
retained classical data vectors as a combination of those in Definitions \ref{def1} and \ref{def3} (for the standard multinomial regression model) and Definitions \ref{def2} \textcolor{red}{and \ref{def4}} (for the OvR regression model), similar statements to Proposition \ref{prop:NewProp} about the existence of the MLEs $\hat{\bbeta}^{\tMM}=\arg\max_{\bbeta} \log L_{\tMM} (\bs; \bbeta)$ and $\hat{\bbeta}^{\tMO}=\arg\max_{\bbeta} \log L_{\tMO} (\bs; \bbeta)$ can be made. 
For example, if no full-histogram MLE exists ($\hat{\bbeta}^{\tSM}$ or $\hat{\bbeta}^{\tSO}$) then no mixture-likelihood MLE exists ($\hat{\bbeta}^{\tMM}$ or $\hat{\bbeta}^{\tMO}$) and no standard likelihood MLE ($\hat{\bbeta}^{\tM}$ or $\hat{\bbeta}^{\tO}$) exists. That is, the standard MLE can't exist without the mixture-likelihood MLE existing, which can't itself exist without the full-histogram MLE existing.
However, the full-histogram model MLE can exist without the mixture-likelihood MLE existing, and the mixture-likelihood MLE can exist without the standard MLE existing.

The choice of  $\tau_k$, for all $k \in \Omega,$ controls the tradeoff between computational efficiency and information loss. On one hand, if $\tau_k$ is too large then we face the original issue of having a huge number of  terms slowing down evaluation of the likelihood function. On the other hand if $\tau_k$ is too low then we risk a loss of efficiency (and perhaps higher computation) compared to higher $\tau_k$.
As a result, one option is to set $\tau_k$ to be the value such that evaluating the integrated bin likelihood term over
over a 
bin $\bb_k$ is less computationally expensive than evaluating the standard likelihood contribution $\tau_k$ times. Some strategies along these lines are explored in the simulations in Section
\ref{sec:sim}.

\subsection{Composite likelihoods for logistic regression models}
\label{sec:comp_improv}

Mixing histogram and micro data can lead to substantial statistical efficiency improvements,
but it does not address the issue of grid-based multivariate histograms becoming highly inefficient as data summaries as the number of covariates ($D$) increases. In particular, the integrals required to compute the likelihood function $L_{SM}(\bs;\bbeta)$ \eqref{eq:SM} have no analytical solution when the outcome has more than two possible classes ($K>2$) and there are more than two explanatory variables ($D>2$). 
Similarly the  integrals in the likelihood function $L_{SO}(\bs;\bbeta)$ 
\eqref{eq:SO} have no analytical solution when more than two explanatory variables are considered ($D>2$). In all non-trivial settings, then, these integrals must be computed numerically. This
can be computationally costly when $D$ is large, which can then defeat the purpose (i.e.~improved computational efficiency) of using data aggregates.

To circumvent the issue of computing the probabilities of data falling in high-dimensional bins, \citet{whitaker2020} proposed 
implementing a  composite likelihood approach. 
This consisted of approximating the likelihood function of a high-dimensional histogram by the weighted product of likelihood functions for lower-dimensional {\em marginal} histograms, which yielded asymptotically consistent likelihood-based parameter estimates \citep{lindsay1988, varin+rf11}. Assuming all weights are equal for simplicity, a $j$-wise composite likelihood function can be expressed as $L^{(j)}(\theta) \propto \prod_{i=1}^m L_i(\theta)$, where $L_i(\theta)$ is the likelihood function of one of $m$ $j$-wise marginal events for a given parameter vector $\theta$. 
In the current context, $L_i(\theta)$ corresponds to a likelihood contribution based on the subset of covariates represented by a $j$-dimensional marginal histogram derived from the full $D$-dimensional histogram, $\theta = \bbeta$ and $m = {D+1 \choose j}$.

However, omitting an important variable in probit and logistic regression analyses will depress the estimated vector of the remaining coefficients
towards zero \citep{Wooldridge2002, cramer2007}. It is therefore non-viable to directly apply a composite likelihood approach to logistic regression problems. However, in the OvR setting and under the assumption that all predictors are independent,
\citet{cramer2007} showed that the non-omitted coefficients of a logistic regression can be written as functions of the regression coefficients in the scenario that no regressor is omitted. This result was primarily aimed at highlighting the effect of omitting variables in a regression analysis context, and had no practical use for e.g.~compensating for zero-depressed parameter estimates, since the established correspondences required information about the variances of the omitted variables, which were unavailable. However, in the composite likelihood setting such information about each covariate is available, and the result of \citet{cramer2007} can therefore be implemented within each marginal likelihood contribution to compensate for the covariates that are omitted in that term. 
We implement this concept in Proposition \ref{prop:composite} below.

In the remainder of this section we construct composite likelihoods for the OvR and histogram-based OvR logistic regression model \cite[the results of][do not hold for multinomial logistic regression]{cramer2007}.
Let $\bi = (i_1,\ldots,i_I)\subseteq \{1,\ldots,D\}$, where for convenience $i_1<\ldots<i_I$, and define by ${\mathcal I}_j=\{\bi: |\bi|=j\}$ the set of all $j$-dimensional subsets of $\{1,\ldots,D\}$. 
We adopt the notation that that a vector with superscript $\bi$  denotes the subvector containing those elements corresponding to the index set $\bi$. For matrices with the superscript $\bi$, the operation is replicated column-wise. E.g.~for $\bi \in {\mathcal I}_j$, $\bX^{(k)\bi} = (X_1^{(k)\bi}, \ldots, X_{N_k}^{(k)\bi}) \in \real^{j \times N_k}$ where $X_n^{(k)\bi} \in \real^j$ is a subvector of $X_n^{(k)}$, $n=1,\ldots,N_k$.
Then if $S_{\bb^{\bi}_k}^{\bi}$ is the random number of observed data points in $\bX^{(k)\bi}$ that fall in bin $\bb^{\bi}_k$, we may construct an $I$-dimensional random {\em marginal} histogram $\bS^{\bi}_k = (S_{\bOne^{\bi}_k}^{\bi},\ldots,S_{\bB^{\bi}_k}^{\bi})$ as the associated vector of random counts from the first bin $\bOne^{\bi}_k = (1,\ldots,1)$ to the last bin $\bB^{\bi}_k = (B^{i_1}_k,\ldots,B^{i_I}_k)$.
The vector $\bS^{\bi}_k$ has length $B^{i_1}_k\times\ldots\times B^{i_I}_k$ and satisfies $\sum_{\bb^{\bi}_k}S_{\bb^{\bi}_k}^{\bi}=N_k$.

The following proposition establishes how to perform approximate composite likelihood estimation for the OvR and histogram-based OvR regressions models using the results in \cite{cramer2007}.
 As independence between predictors, as assumed by \cite{cramer2007}, is unrealistic, the Proposition also extends the results of \citet{cramer2007} to account for the correlation between the included set of predictor variables within each composite likelihood term and the omitted variables. 
Without loss of generality, consider a random vector $X \in \real^D$ and for $\bi \in {\mathcal I}_j$, let $X^{\bi} \in \real^j$ represent the observed variables of $X$. 
Further let $\mathcal{I}_1^{-\bi} = \{1,\ldots,D\}\backslash\{\bi\}$
such that for all $i' \in \mathcal{I}_1^{-\bi}$, $X^{i'}$ represents an omitted variable of $X$. 
Following \cite{cramer2007} we define the omitted variables $X^{i'}$ to be a linear function of the observed variables via $X^{i'} = \alpha_{\bi i'}^\top X^{\bi} + \epsilon_{\bi i'}$, where $\alpha_{\bi i'} = (\alpha_{i_1 i'}, \ldots, \alpha_{i_j i'})^\top \in \real^j$ and $\epsilon_{\bi i'}\sim N(0,\lambda_{\bi i'}^2)$.
Denote $\Cov(\epsilon_{\bi i_1'}, \epsilon_{\bi i_2'}) = \lambda_{\bi i_1' i_2'}$.

\begin{prop}
\label{prop:composite}
The $j$-wise approximate composite likelihood functions for the standard  and the histogram-based $D$-dimensional  OvR logistic regression models are respectively given by
$$
L_{\tO}^{(j)}(\bx, \textcolor{red}{\by}; \bbeta) = \prod_{\bi \in \mathcal{I}_j} L_{\tO} (\bx^{\bi}, \textcolor{red}{\by}; \widetilde{\bbeta}^{\bi})
\quad \textrm{and} \quad
L_{\tSO}^{(j)}(\bs; \bbeta) = \prod_{\bi \in \mathcal{I}_j} L_{\tSO} (\bs^{\bi}, \textcolor{red}{\by}; \widetilde{\bbeta}^{\bi}),
$$
where the lower dimensional regression observed coefficients are given by $\widetilde{\bbeta}^{\bi} = \left( \widetilde{\beta}^{\bi}_1, \ldots, \widetilde{\beta}^{\bi}_K \right) \in \real^{(j+1) \times K}$ where
\begin{align}
\label{eq:beta_tilde}
\widetilde{\beta}^{\bi}_k = \frac{ {\beta}^{\bi}_k + 
\left[ 0, \left( \sum_{i' \in \mathcal{I}_1^{-\bi}} \beta_{k}^{i'} \alpha_{\bi i'}\right)^\top \right]^\top }
{\sqrt{
1 + \frac{\pi^2}{3}\sum_{i_1' \in \mathcal{I}_1^{-\bi}} \left[ (\beta_{k}^{i_1'} \lambda_{\bi i_1'})^2 + 2  \sum_{i_2' \in \mathcal{I}_1^{-\bi}, i_2' \neq i_1'} \beta_{k}^{i_1'} \beta_{k}^{i_2'} \lambda_{\bi i_1' i_2'} \right]
}
}\in \real^{(j+1)}.
\end{align}
If there is neither complete nor quasi-complete separation in the full $D$-dimensional dataset $\bx$ (Definition \ref{def1}) for any binary logistic model, then $L_{\tO}^{(j)}(\bx, y; \bbeta)$ will have a unique global maxima. Similarly, if there is neither complete nor quasi-complete separation in any of the sets of marginal histograms $\bs^{\bi}$, $\bi\in \textcolor{red}{\mathcal{I}_j}$, then $L_{\tSO}^{(j)}(\bs; \bbeta)$ will have a unique global maxima. See Appendix \ref{proofprop:composite} for a derivation and proof.
\end{prop}

Note that $L_{\tO}^{(j)}(\bx, y; \bbeta)$ and $L_{\tSO}^{(j)}(\bs; \bbeta)$ are {\em approximate} composite likelihood functions rather than true composite likelihood functions. The resulting maximum composite likelihood estimators ($\hat{\bbeta}^{(j)}_{\tO}$ and $\hat{\bbeta}^{(j)}_{\tSO}$) are not unbiased or consistent. They are, however, reasonable parameter estimates if one is motivated to estimate logistic regression model parameters within the composite likelihood framework (as is the case here), that are more accurate than those estimated via a naive composite likelihood implementation (which we define here as simply $\tilde{L}_{\tSO}^{(j)}(\bs; \bbeta) = \prod_{\bi \in \mathcal{I}_j} L_{\tSO} (\bs^{\bi}, \by; \bbeta^{\bi})$). We numerically demonstrate the performance of these estimators against the naive implementation in the simulation study in Section \ref{sim2}.
However, when our primary aim is model predictive accuracy, we will demonstrate that the performance of the fitted model using the approximate composite likelihood MLE ($\hat{\bbeta}^{(j)}_{\tO}$ or $\hat{\bbeta}^{(j)}_{\tSO}$) is highly competitive with using the full-data standard MLE, $\hat{\bbeta}_{\tO}$, while, in the case of $\hat{\bbeta}^{(j)}_{\tSO}$, being much more computationally efficient to obtain.

Following similar arguments to before, it is clear that $L_\tSO^{(j)}(\bx, \by; \bbeta) \rightarrow L_\tO^{(j)}(\bs; \bbeta)$ as the number of histogram bins becomes large while the volume of each bin approaches zero.
\citet{whitaker2020} proved the asymptotic normality and consistency of the histogram-based (true) composite likelihood estimator, and highlighted that parameter variance consistency requires
the number of bins in each histogram to become large and the number of histograms representing the data be close to N, the number of data points. While these results are not directly applicable under the approximate composite likelihood of Proposition \ref{prop:composite}, we intuitively expect that that the estimated variances of $\hat{\bbeta}^{(j)}_{\tO}$ and $\hat{\bbeta}^{(j)}_{\tSO}$ will be inflated compared to that of $\hat{\bbeta}_{\tO}$ unless the same conditions hold.

The estimation procedure for the histogram-based $D$-dimensional OvR logistic regression model is presented in Algorithm~\ref{alg:alg_SO}. It illustrates the use of the $j$-wise approximate composite likelihood given in Proposition~\ref{prop:composite}, and considers the full likelihood equivalent of Proposition~\ref{prop1} as the special case of $j=D$. This algorithm can be adapted for the multinomial logistic regression model by considering the likelihood function $L_{\tSM} (\bs; \bbeta)$ given in \eqref{eq:SM} rather than $L_{\tSO} (\bs; \bbeta)$, but only for $j=D$. Estimation for the mixture models $\tMM$ and $\tMO$ is achieved in a similar manner. 
\begin{algorithm}
	\caption{Estimation procedure using the $j$-wise approximate composite likelihood for the histogram-based $D$-dimensional OvR logistic regression model.}
	\label{alg:alg_SO}
	\textbf{Input data:} $Y \in \{1, \ldots, K\}^N, \bX \in \real^{N\times D}$ \;
	\nextnr \textbf{Divide} $\bX$ into $\bX^{(1)}, \ldots, \bX^{(K)}$ where $\bX^{(k)} = \left( X_n | Y_n =k, n=1,\ldots,N \right) \in \real^{D\times N_k}$\;
	\nextnr Define the locations $\Upsilon_{\bb_k}$ of the bins $\bb_k$\;
		\nextnr \eIf{j = D}{
			\nextnr Construct $\bS = (S_1, \ldots, S_K)$ by applying \eqref{eq:hist_const} to $\bX^{(1)}, \ldots, \bX^{(K)}$\;
			\nextnr Evaluate $ L_{\tSO} (\bs; \bbeta)$ using \eqref{eq:SO} \;
			\nextnr Compute $\hat{\bbeta}^{\tSO}=\arg\max_{\bbeta} \log L_{\tSO} (\bs; \bbeta)$ \;
		}{
			\nextnr \For{$\bi \in {\mathcal I}_j$}{
				\nextnr Construct  $\bS^{\bi} = (\bS_1^{\bi}, \ldots, \bS_k^{\bi})$ by applying  \eqref{eq:hist_const} to subvectors $\bX^{(1)\bi}, \ldots, \bX^{(K)\bi}$ \;
				\nextnr Estimate $\alpha_{\bi i_1'}, \lambda_{\bi i_1'}$ and $\lambda_{\bi i_1' i_2'}$ for all $i_1', i_2' \in \mathcal{I}_1^{-\bi}, i_1' \neq i_2'$ \;
				\nextnr Evaluate $\widetilde{\bbeta}^{\bi}$ using \eqref{eq:beta_tilde} and  $ L_{\tSO} (\bs^{\bi}, y; \widetilde{\bbeta}^{\bi})$ using  \eqref{eq:SO} \;
			}
			\nextnr Compute $\hat{\bbeta}^{(j)}_{\tSO} = \arg\max_{\bbeta} \sum_{\bi \in \mathcal{I}_j} \log L_{\tSO} (\bs^{\bi}, y; \widetilde{\bbeta}^{\bi})$
}		
\end{algorithm}

In specific cases we can obtain a closed-form approximate composite likelihood function for a $D$-dimensional random histogram, $L_{\tSO}(\bs;\bbeta)$. In the particular case of a binary outcome ($K=2$) and using $j=1$ to construct the composite likelihood from all univariate marginal events 
(the set $\mathcal{I}_1=\{1,\ldots,D\}$),
we have
\begin{align}
\label{eq:SO1}
L_{\tSO}^{(1)}(\bs; \bbeta)= 
\prod_{i \in \mathcal{I}_1} \prod_{k=1}^2 \prod_{b_{i_k} = 1}^{B_k^i}
\left[
\left( \frac{1}{\widetilde{\beta}_{k1}^i} \right)^2
\log \left( \frac{1+ e^{\widetilde{\beta}_{k0}^i + \widetilde{\beta}_{k1}^i y_{b_i}^i } }{1 + e^{\widetilde{\beta}_{k0}^i + \widetilde{\beta}_{k1}^i y_{b_i-1}^i }} \right)
\log \left( \frac{1+ e^{-\widetilde{\beta}_{k0}^i - \widetilde{\beta}_{k1}^i y_{b_i-1}^i } }{1 + e^{-\widetilde{\beta}_{k0}^i - \widetilde{\beta}_{k1}^i y_{b_i}^i }} \right)
\right]^{s_{b_k}^i},
\end{align}
where $\widetilde{\beta}^i_k = (\widetilde{\beta}^i_{k0}, \widetilde{\beta}^i_{k1}) \in \real^2$ and the bin indexed by $\bb^i_k = b_{i_k}$ is constructed over the space $\Upsilon^i_{\bb^i_k} = (y^i_{b_i-1}, y^i_{b_i}]$. 
In all other cases, the integrals in $L_{\tSO}(\bs;\bbeta)$ \eqref{eq:SO} require numerical estimation.

Evaluating \eqref{eq:beta_tilde} within the approximate composite likelihood requires knowledge of $\alpha_{\bi i_1'}, \lambda_{\bi i_1'}$ and $\lambda_{\bi i_1' i_2'}$ for all $i_1', i_2' \in \mathcal{I}_1^{-\bi}, i_1' \neq i_2'$. The $\alpha_{\bi i_1'}$ terms are the coefficients explaining the variations of an unobserved variable $X^{i_1'}$ as a linear function of the observed variables $X^{\bi}$ (i.e.~$X^{i'} = \alpha_{\bi i'}^\top X^{\bi} + \epsilon_{\bi i'}$). In the case where $\bi =i \in \mathcal{I}_1$ (i.e.~a simple linear regression with $j=1$ as in \eqref{eq:SO1}), 
an estimate of $\alpha_{i i_1'}$ is $\Cov(X^i, X^{i_1'}) / \Var(X^i)$. In this context, rewriting $\epsilon_{i i_1'}=X^{i_1'} - \alpha_{i i_1'} X^i$, we also have that 
\begin{align*}
\hat{\lambda}_{i i_1'}^2 = \Var(X^{i_1'}) - \frac{\Cov(X^i, X^{i_1'})^2}{\Var(X^i)}
\quad \textrm{and} \quad 
\hat{\lambda}_{i i_1' i_2'} = \Cov(X^{i_1'}, X^{i_2'}) - \frac{\Cov(X^i, X^{i_1'}) \Cov(X^i, X^{i_2'}) }{\Var(X^i)}.
\end{align*}
Knowledge of variances and covariances between the covariates $X$ is similarly required
when two or more predictors are considered in each approximate composite likelihood contribution i.e.~when $\bi \in \mathcal{I}_j,$ for $j\geq 2$. Ideally these variances and covariances should be computed and stored prior to the data aggregation process i.e.~on the full dataset, but if this information is unavailable, variance and covariance estimates can be derived directly from the histograms either with \citep{beranger+ls18} or without \citep{billard+d03, billard2011} parametric assumptions.
Clearly, the assumption that a missing variable can be written as a linear combination of observed variables may not hold. Where viable, transformations (e.g.~Box-Cox) 
or more flexible regression models 
can be applied to provide a more realistic model, either regressing on the transformed covariates or modifying the 
form of $\tilde{\beta}^{\bi}_k$ \eqref{eq:beta_tilde} as required.

Finally, suppose that we again consider the case $j=1$ so that the approximate composite likelihood is constructed with each term comprising each covariate separately \eqref{eq:SO1}. This is the most computationally efficient histogram-based likelihood as it is based solely on univariate histograms (and known covariances with the other covariates). 
This construction implies that only univariate marginal histograms are required.
\citet{beranger+ls18} introduced two likelihood constructions for histogram-valued variables. The first, which we have used until now, assumes that histogram bins are fixed and the corresponding counts are random, which works straightforwardly in $D$-dimensions. The second  construction is a quantile-based approach for univariate variables only, where the bin locations are assumed random and the counts fixed. Defining bin locations using quantiles can better describe the behaviour of the underlying data, and also has the advantage of retaining some of the micro-data (at the observed quantiles) which resembles the mixture of histogram and standard likelihood approach of Section~\ref{sec:stat_improv}.

In this setting, for each $k \in \Omega$ and each marginal component $i \in \mathcal{I}_1$, define a vector of order statistics $t = (t_1,\ldots, t_B)^\top$ where $1 \leq t_1 \leq \ldots \leq t_B \leq N_k$, such that a quantile-based histogram-valued random variable is obtained through the aggregation function
\begin{align}
\label{eq:hist_const_os}
S_k^i = \dot{\pi}(\bX^{(k)i}): \real & \rightarrow \mathcal{S} = \{ (a_1, \ldots, a_B) \in \real^B: a_1 < \ldots < a_B \} \times \mathbb{N} \\
\bx^{(k)i} &\mapsto s_k^i = \left( \bx^{(k)i}_{(t_1)}, \ldots, \bx^{(k)i}_{(t_B)}, N_k \right ), \nonumber
\end{align}
where $\bx^{(k)i}_{(t_b)}$ denotes the $t_b$-th order statistic of $\bx^{(k)i} \in \real$. (Note that to ease notation we have omitted superscripts and subscripts related to $i$ and $k$ in the order statistics $t$.)
The $b$-th histogram bin is then defined over the range $(s^i_{kb-1}, s^i_{kb}]$ with fixed counts of underlying datapoints $t_b - t_{b-1}$, for $b = 1,\ldots,B+1$, where $s_0 = -\infty, s_{B+1} = + \infty$, $t_0 = 0$ and $t_{B+1} = N_k + 1$.
If each covariate is aggregated via \eqref{eq:hist_const_os}, then the resulting approximate composite likelihood function is 
$$
L_{\tOO}^{(1)} (\bs; \bbeta) = L_{\tO}^{(1)}(\{\bx^{(k)i}\}; \bbeta)L_{\tSO}^{(1)}(\bs; \bbeta),
$$
where $\bs = (s_1, \ldots, s_K)$ and $s_k = (s_k^1, \ldots, s_k^D)$ with $s_k^i$ defined in \eqref{eq:hist_const_os}, and where $L_{\tSO}^{(1)}$ is the likelihood shown in (\ref{eq:SO1}).

\section{Simulation studies}
\label{sec:sim}

We now examine the parameter estimation and classification capabilities of the methods developed in Section \ref{sec:sym} based on simulated data. We consider both the statistical and computational performance of the histogram-based analyses compared to  the standard full-data approach.

In the following we set the number of possible outcomes $K$ to define $\Omega = \{1, \ldots, K\}$, the domain of the response variable $Y$. We obtain the $(D \times 2N)$ matrix of covariates $\bX$ by generating $2N$ observations from a specified $D$-dimensional distribution. Given a fixed matrix of regression coefficients $\bbeta \in \real^{(D+1)\times K}$,  for each $n=1, \ldots, 2N$ we compute the probability of every outcome in $\Omega$ using \eqref{eq:logodds}. These probabilities are then used to generate $\bY \in \Omega^{2N}$ from a multinomial distribution. The dataset $(\bY, \bX)$ is then split into equally sized training and test datasets, and the estimates $\hat{\bbeta}^{\tM}, \hat{\bbeta}^{\tO}, \hat{\bbeta}^{\tMM}, \hat{\bbeta}^{\tO(j)}$ and $\hat{\bbeta}^{\tSO(j)}$ are obtained by maximising their respective likelihood functions on the training dataset. Using the test dataset we compute the prediction accuracy (PA) of a model (and estimation procedure) as
$$
PA = \frac{1}{N} \sum_{n=1}^N \boldsymbol{1} \{ Y_n^{\Pred} = Y_n \},
$$
where $Y_n^{\Pred} = \argmax_{k \in \Omega} P (Y=k | X=X_n), n=1, \ldots, N$, 
denotes the predicted outcome under the model (multinomial or OvR) with estimated coefficients $\hat{\bbeta}^{\tM}, \hat{\bbeta}^{\tO}, \hat{\bbeta}^{\tMM}, \hat{\bbeta}^{\tO(j)}$ or $\hat{\bbeta}^{\tSO(j)}$. 
After replicating the above $1\,000$ times, we report the mean squared errors (MSE) and mean prediction accuracies of each estimator.

In Section~\ref{sec:stat_improv} we proposed to improve the statistical and computational efficiency of the MLE by 
retaining the data underlying a histogram bin, rather than using the bin itself, if the number of underlying observations in the bin 
$s_{\bb_k}$ is less than $\tau_k$. 
We use quadrature methods to perform the integrations in \eqref{eq:SM},
or which the number of function evaluations used to approximate the integral needs to be specified. 
We specified a value for which adding more function evaluations 
only yielded negligible changes in the estimated value of the integral. (A similar approaches could be adopted to determine the most appropriate number of marginal bins $B$.) 
Through experimentation we determined that using $2^{j}$ function evaluations for each integral globally produced small enough approximation errors when integrating over $j$-dimensional bins to obtain comparable results to the classical model. As this implies the minimum number of evaluations necessary for a reasonable approximation of the integrals across all bins,
we set $\tau_k =  2^j$.

\subsection{Varying the number of bins, $B$}
\label{sec:sim1}

We specify (training and test) datasets each comprising $N=25\,000$ observations for which the response variable $Y$ can take values in $\Omega = \{ 1, 2, 3 \}$ ($K=3$) conditional on $D=5$ explanatory variables.
The true vector of regression coefficients $\bbeta_{true}$ has entries randomly drawn from a 
$U[-5,5]$ distribution. The explanatory variables are drawn from $D$-dimensional normal and skew-normal 
distributions, with correlation matrices containing zero correlations (the identity matrix) or correlations drawn from $U[-0.75,0.75]$. The elements of the skew-normal slant vector are drawn from $U[-7,7]$. 
While the correlation parameter of the skew-normal distribution is not equivalent to the correlation matrix of the associated random variable, 
skew-normal data simulated using the identity matrix as the correlation parameter typically have low correlations.
When aggregating the design matrix $\bX$ into a histogram through \eqref{eq:hist_const}, an equal number $B$ of bins is set for each margin and each outcome $k$, i.e.~$\bB_k = (B,B,B,B,B,B)$ for all $k \in \Omega$.

We use both the full multinomial regression (M) model fit using \eqref{eq:MLR} and the OvR (O) model fit using the full likelihood \eqref{eq:OVR} as reference fits. 
We also fit the multinomial mixture (MM) model of histogram and underlying classical data \eqref{eqn:MMlik},
and the univariate approximate composite likelihood $L_\tO^{(1)}$ (see Proposition~\ref{prop:composite}). 
For the histogram-based, univariate approximate composite likelihood 
\eqref{eq:SO1} we make the assumption that the covariates are either independent ($\alpha_{\bi i'}=0$ and $\lambda_{\bi i_1' i_2'} = 0$ in \eqref{eq:beta_tilde}) or that the variance-covariances of the covariates are to be estimated.

\begin{figure}[t!]
\centering
\includegraphics[width=0.44\textwidth]{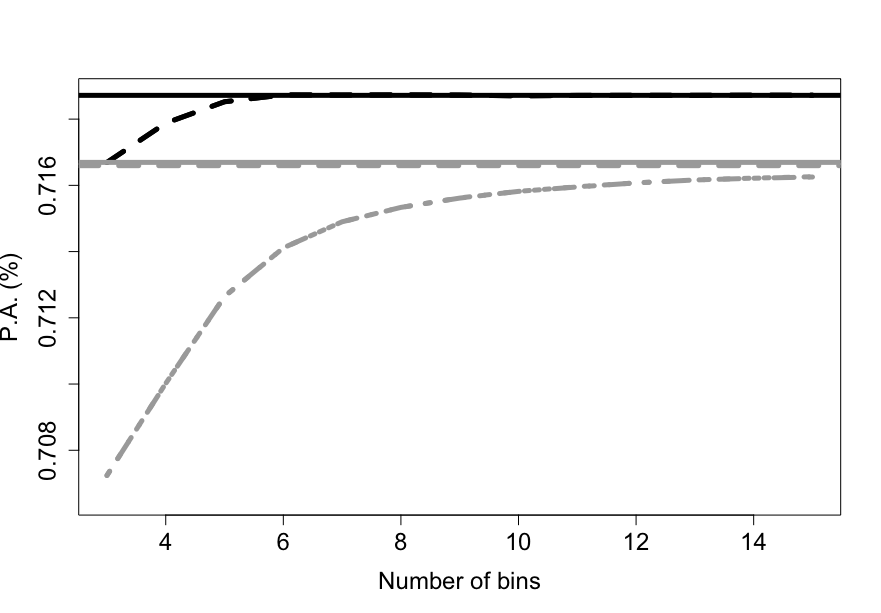}
\includegraphics[width=0.44\textwidth]{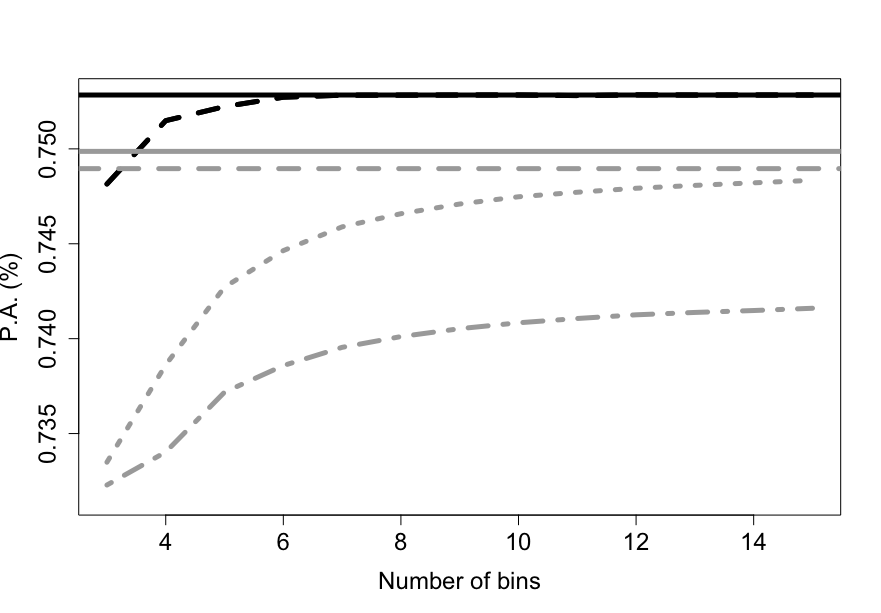}
\includegraphics[width=0.44\textwidth]{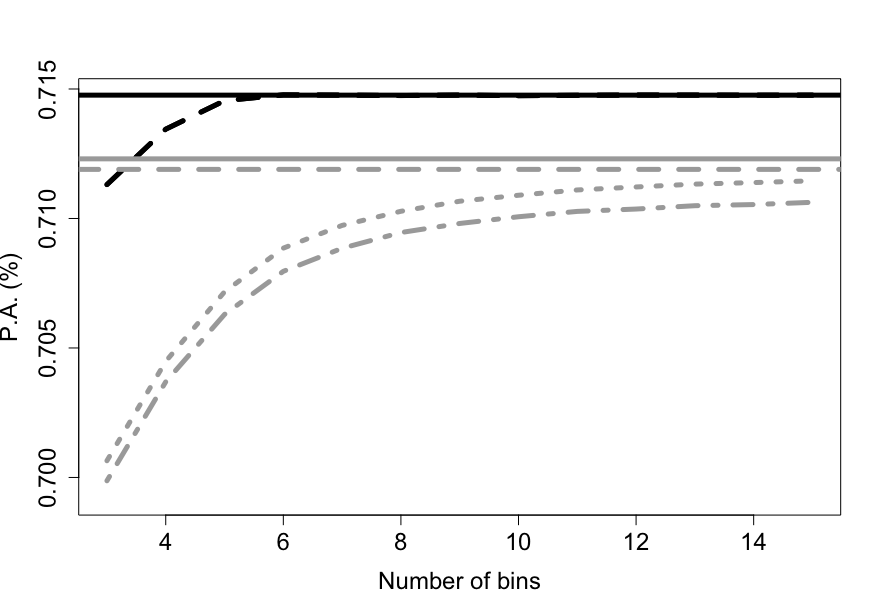}
\includegraphics[width=0.44\textwidth]{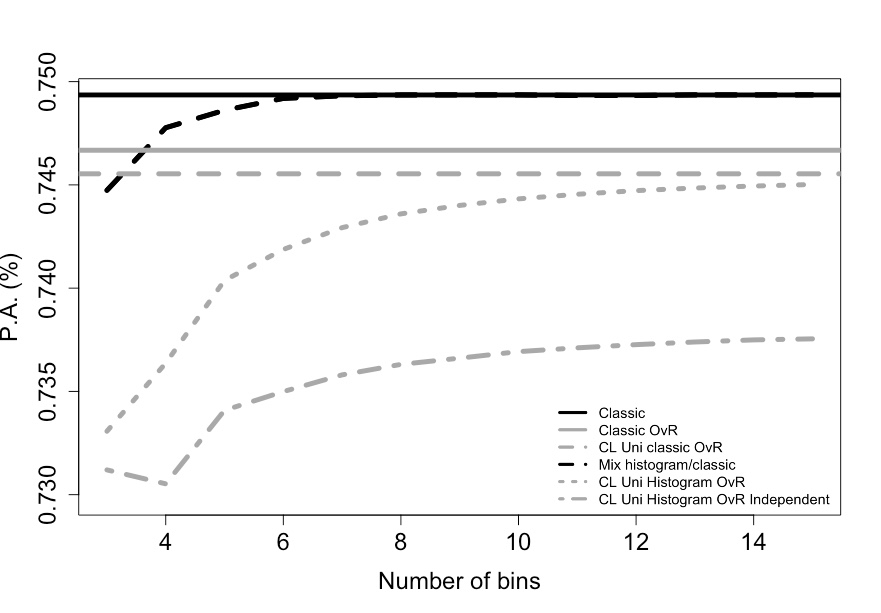}
 \caption{Average prediction accuracy (PA) computed over 1\,000 replications for the multinomial model using full likelihood (solid black), mixture multinomial model (dashed black), OvR model using full likelihood (solid grey) and approximate composite likelihood (dashed grey), histogram-based OvR model using approximate composite likelihood assuming independence of the covariates (dot dashed grey) and using additional covariate assumption (dotted grey). Top panels consider covariates simulated from the multivariate normal distribution and bottom panels using the skew normal distribution. Left panels assume the covariates have zero correlation parameter, and right panels use non-zero correlations.
}
\label{fig:sim1_pred}
\end{figure} 
Figure~\ref{fig:sim1_pred} illustrates the mean prediction accuracies (over $1\,000$ replicates) as a function of the number of bins $B$, obtained for each of the above models and estimation procedures. 
For the full dataset, using an approximate composite likelihood approach to fit the OvR model (dashed grey line) yields comparable prediction accuracies to the full likelihood approach (solid grey line), in particular when the covariates are independent of each other (left panels).

When the empirical variance-covariance matrix is used, the histogram-based OvR model fitted using the $L_{\tSO}^{(1)}$ approximate composite likelihood (grey dotted line) is able to obtain comparable prediction accuracies to the $L_{\tO}^{(1)}$ full-data approximate composite likelihood OvR model (solid grey line), for a reasonable ($\approx 10$ marginal bins) level of data aggregation, and for any covariate distribution (rows). This clearly demonstrates that $L_{\tSO}^{(j)}(\bx,\by;\bbeta) \rightarrow L_{\tO}^{(j)}(\bs;\bbeta)$ as discussed in Section \ref{sec:comp_improv}.
It also performs well compared to the analysis on the full data (solid black line) when the covariates are independent (left panels; note the small $y$-axis scale).

When the empirical variance-covariance matrix is unavailable and the correlations between explanatory variables are assumed to be independent ($\alpha_{\bi i'}=0$ and $\lambda_{\bi i_1' i_2'} = 0$), the histogram-based approximate composite likelihood model (dot-dashed grey line) still produces reasonable prediction accuracies.
However, as should be expected, there is a clear loss in performance compared to when the covariances are known, for densities with high covariate correlations (right panels)
and asymmetric distributions (bottom panels).

The prediction accuracies obtained from the multinomial model (dashed black line) have converged relatively quickly to its classical equivalent (solid black line) requiring only around $B=6$ bins per margins. 
To understand this, note that the histogram comprises $6^5 = 7\,776$ bins for $25\,000$ raw data points and $3$ possible outcomes. As a result there will be numerous empty bins, and bins with low covariate vector counts. In addition, because $\tau_k = 2^5=32$ (see above), data are only aggregated if there are more than 32 observations in each bin. Consequently, for increasing $B$, the summarised dataset can quickly retain the most
important information from the original dataset.
The multinomial model gives higher overall prediction accuracies than the OvR model in each case, however recall that the data were simulated according to the multinomial model, giving it a natural advantage. While the multinomial model is generally preferred over the OvR model here, in practice the One-vs-Rest approach can outperform the multinomial model for some datasets, and can give almost as good results in many other cases \citep[e.g.][]{eichelberger2013}.

\begin{figure}[t!]
\centering
\includegraphics[width=0.44\textwidth]{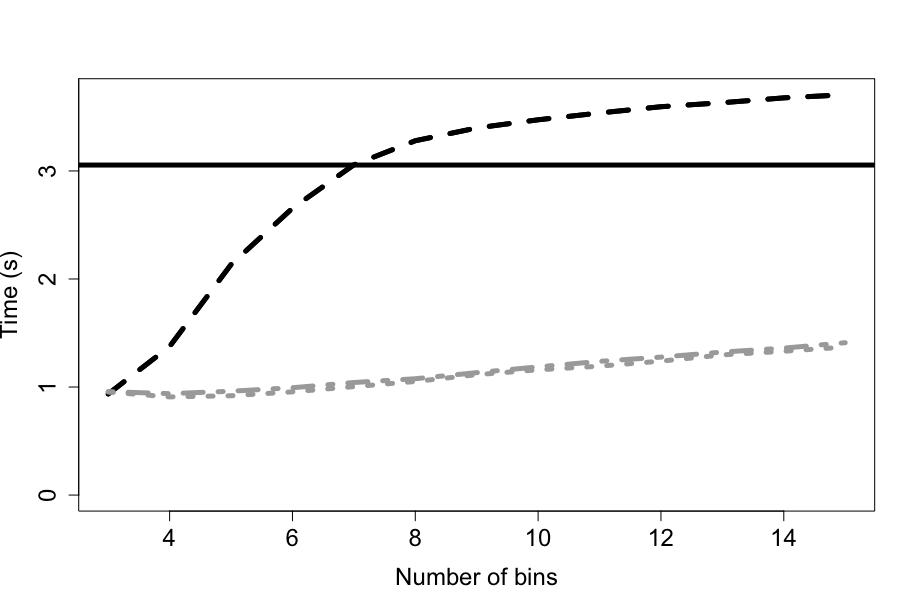}
\includegraphics[width=0.44\textwidth]{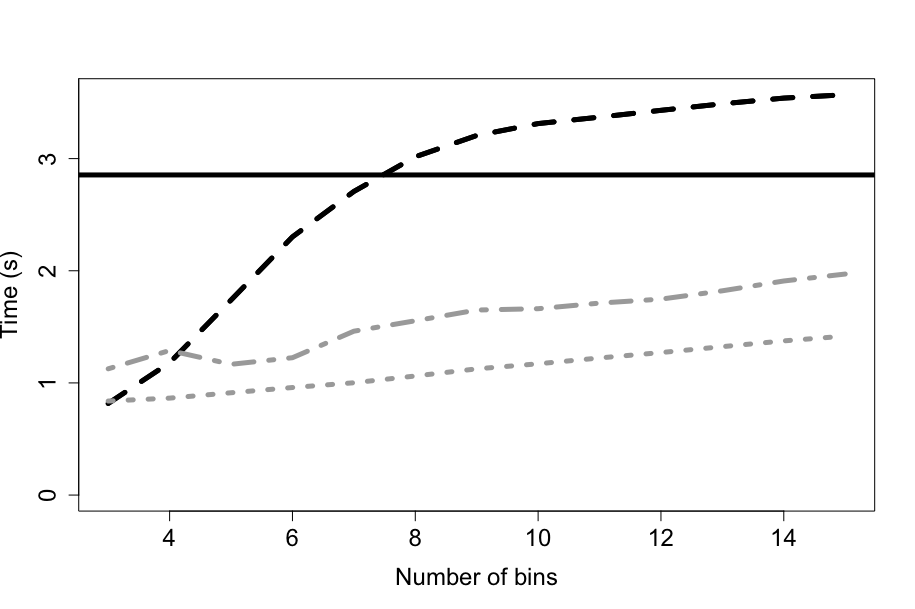}
\includegraphics[width=0.44\textwidth]{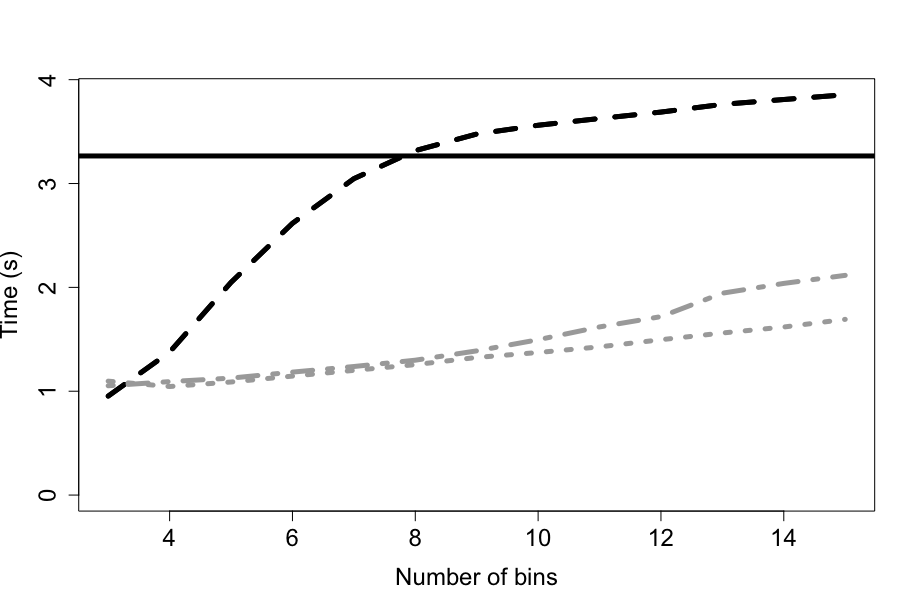}
\includegraphics[width=0.44\textwidth]{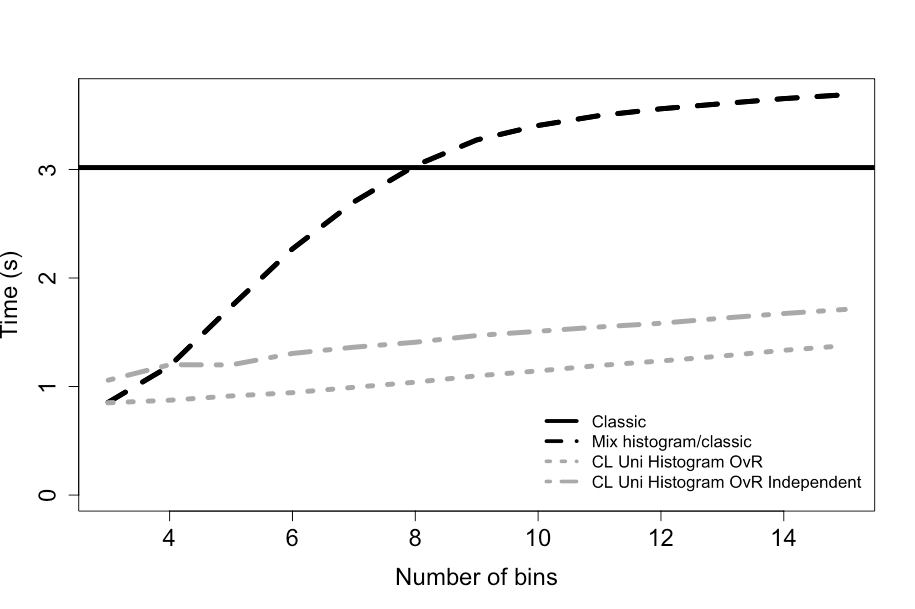}
 \caption{Average computation time (in CPU seconds) over 1\,000 replications for the multinomial model using full likelihood (solid black), mixture multinomial model (dashed black), histogram-based OvR model using approximate composite likelihood assuming independence of the covariates (dot dashed grey) and using additional covariate assumption (dotted grey). Top panels consider covariates simulated from the multivariate normal distribution and  bottom panels using the skew normal distribution. Left panels use covariates from a zero correlation parameter, and right panels use non-zero correlations.
}
\label{fig:sim1_time}
\end{figure} 
Figure~\ref{fig:sim1_time} illustrates the mean computational efficiency of fitting each model. 
It highlights the computational superiority of the histogram-based OvR model when univariate components are included in the likelihood ($L_{\tSO}^{(1)}(\bs;\bbeta)$; grey lines) against the full data multinomial model (black lines). The computation time increases as the number of bins increases when $L_{\tSO}^{(1)}(\bs;\bbeta)$ is used, however comparable predictions to the full multinomial model are achieved for $B\approx 10$ (c.f.~Figure \ref{fig:sim1_pred}) at a significantly cheaper computational cost. 
For the mixture multinomial model (black dashed lines), the computation time increases strongly with increasing $B$. This phenomena is due to the intractability of the integrands in \eqref{eq:SM} requiring numerical integration, and our choice of $\tau_k$. However, for this setup $B\approx 5$ is sufficient to provide comparable predictions to the multinomial model (Figure \ref{fig:sim1_pred}), and with lower computational overheads than the complete data case.

Even though the $L_{\tSO}^{(1)}(\bs;\bbeta)$ and $L_{MM}(\bs;\bbeta)$ likelihood approaches for histogram-valued data increase in computational intensity with increasing $B$, relative to the classical $M$ model, we need to keep in mind that $N=25\,000$ observations are considered. Increasing $N$ will directly increase the computational time of the classical approach (solid black line), while computational overheads will remain relatively unchanged for the histogram-based methods (where computation is proportional to the number of bins, not datapoints within bins). Consequently, there are clear computational benefits to employing a histogram likelihood approach when analysing extremely large datasets. We explore this in Section \ref{sim2}.

\subsection{Varying the number of underlying observations, $N$ and comparison with subsampling}
\label{sim2}

Aggregating data into summaries and performing an analysis on these new ``datapoints'' seems a good strategy when the sample size is large. It is natural to compare this  approach to other popular techniques for downsizing data volume, such as subsampling algorithms. We use the two-step subsampling scheme given by \citet{wang2018} for logistic regression modelling,
which first uniformly draws a subsample of size $r_0=1\,000$ from the dataset to produce a provisional MLE
$\hat{\bbeta}_0$, and uses this to produce an optimal 
weight for each datapoint.
The second step then draws with replacement a subsample of size $r=1\,000$ using the optimal weights, and then determines the final estimate of $\bbeta$ using the total subsample of size $r_0 + r$. 

In the following binary response ($K=2$) experiment each element in the true vector of regression coefficients is generated from $U[-1,1]$, and the number of observations, $N$, 
varies between $5\,000$ and $100\,000$. The explanatory variables are drawn from $8$-dimensional skew-normal distributions ($D=8$), with either zero correlations (identity matrix) or correlations drawn uniformly on $[0,0.75]$. The slant vector of the skew-normal distribution is drawn from $U[-10,10]$. (Recall that the identity correlation matrix for the skew-normal distribution does not lead to independent covariates, but rather low correlations between the covariates.)  
Note that in the case of binary responses the multinomial and OvR models are identical. After aggregating the covariates into histograms with $B=15$ bins for each margin, the histogram-based OvR model is fitted using $L_{\tSO}^{(1)}$ and $L_{\tSO}^{(2)}$ (\textcolor{red}{i.e.}~univariate and bivariate marginal histograms), including covariate correlations. The $L_{\tSO}^{(1)}$ model is also fitted assuming independence between covariates.

\begin{figure}[t!]
\centering
\includegraphics[page=1,width=0.4\textwidth]{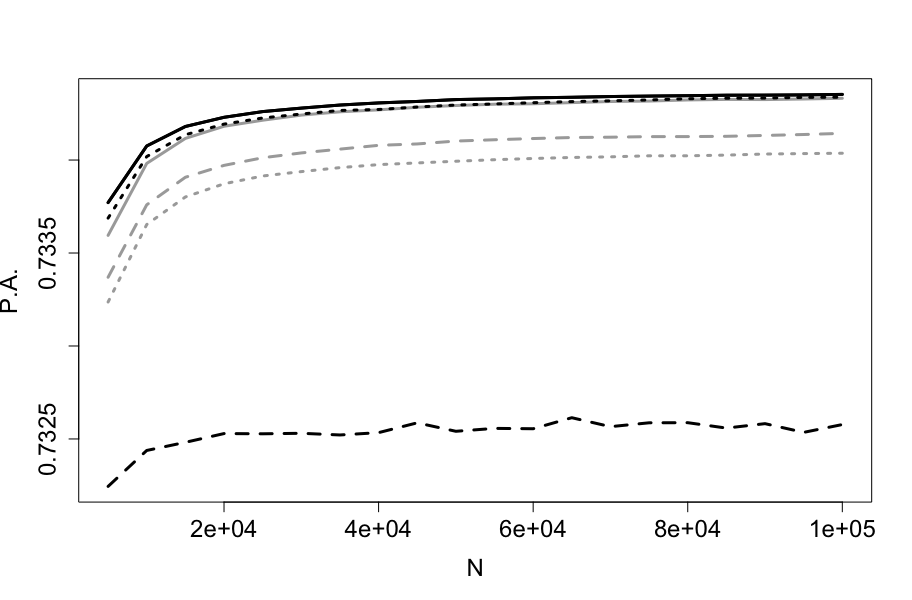}
\includegraphics[page=1,width=0.4\textwidth]{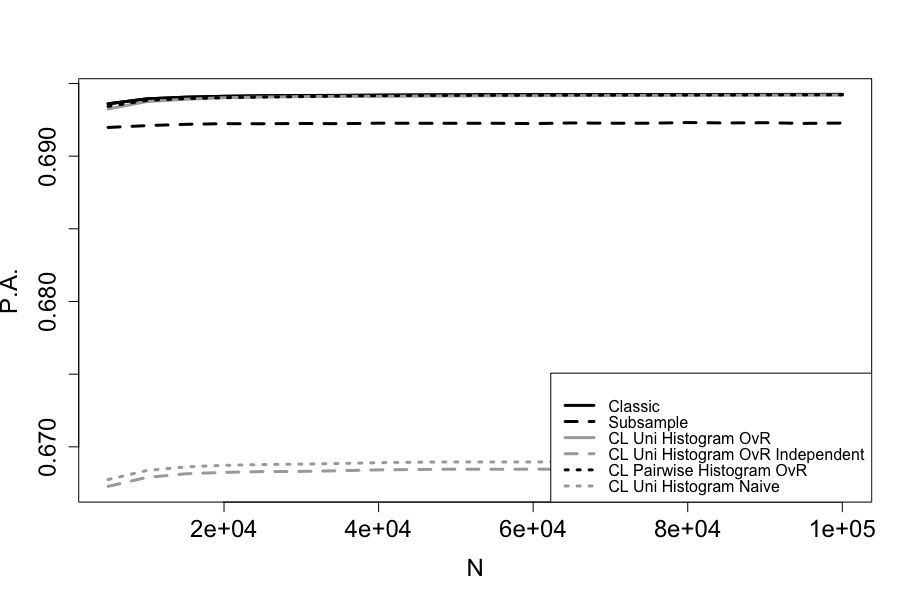}
 \caption{Mean prediction accuracies (P.A.) using the multinomial model on the full data (solid black line), subsampled data (dashed black line) and the histogram-based OvR model using $L_{\tSO}^{(1)}$ with independence assumption (dashed grey line), $L_{\tSO}^{(1)}$ with correlations (solid grey line), $L_{\tSO}^{(2)}$ (dotted black line) and the naive composite likelihood model (dotted grey line) as a function of the number of datapoints $N$. The covariates are generated from $8$-dimensional skew-normal distributions,  with zero correlations (left), or correlations drawn uniformly on $[0,0.75]$ (right). The responses have two possible outcomes ($K=2$). Results are based on $1\,000$ replicate analyses.
 }
\label{fig:sim2_pred}
\end{figure}

Figure~\ref{fig:sim2_pred} shows that the mean prediction accuracies obtained by each method are increasing functions of the sample size $N$. Overall the symbolic based methods yield higher prediction accuracies than the subsampling approach when the covariate correlations are incorporated, and the more informative bivariate histogram setup  $L_{\tSO}^{(2)}$ will outperform the univariate histogram-based $L_{\tSO}^{(1)}$.  When the covariates exhibit low correlations, the $L_{\tSO}^{(1)}$ model provides significant improvements over the naive composite likelihood analysis, regardless of whether correlations are incorporated. When there are correlations between the covariates, the $L_{\tSO}^{(1)}$ model only provides significant improvements over the naive composite likelihood analysis if the covariate correlations are included. The magnitude of the variations in the prediction accuracy confirms that the extra efforts to use $L_{\tSO}^{(2)}$ are not justified in this case, and that univariate marginal histograms provide enough information and produce comparable results to a classical full data analysis. 

\begin{figure}[t!]
\centering
\includegraphics[page=1,width=0.4\textwidth]{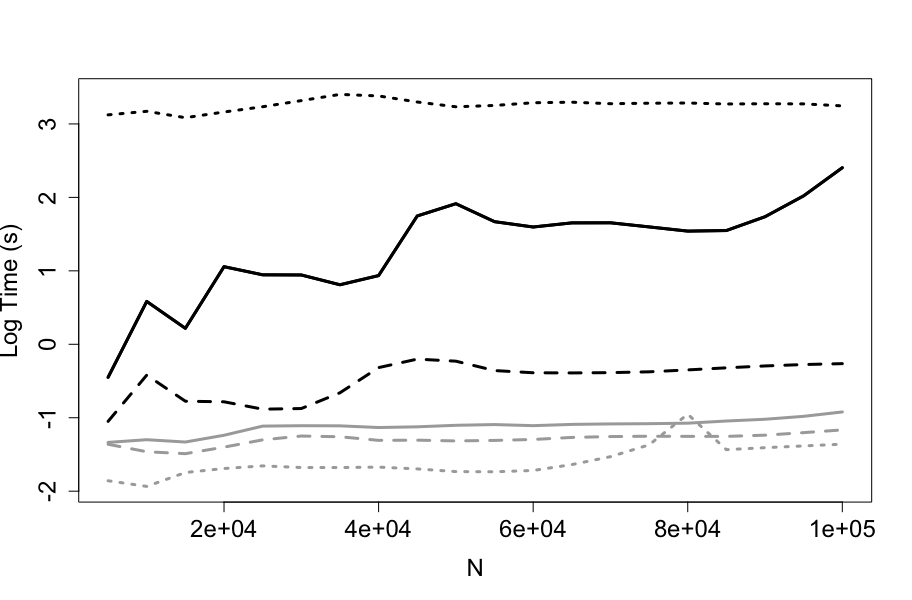} 
\includegraphics[page=1,width=0.4\textwidth]{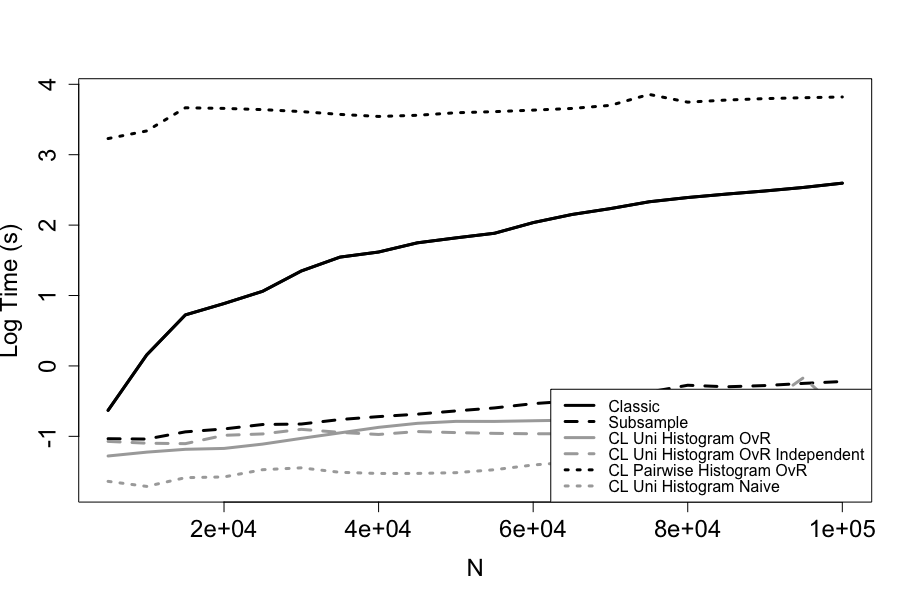}
 \caption{Mean total computation times (in CPU seconds) for the multinomial model on the full data (solid black line), subsampled data (dashed black line) and the histogram-based OvR model using $L_{\tSO}^{(1)}$ with independence assumption (dashed grey line), $L_{\tSO}^{(1)}$ with correlations (solid grey line), $L_{\tSO}^{(2)}$ (dotted black line) and the naive composite likelihood model (dotted grey line) as a function of the number of datapoints $N$. The covariates are generated from $8$-dimensional skew-normal distributions, considering zero (left) and non-zero (right) correlation parameters. Results are based on $1\,000$ replicate analyses.}
\label{fig:sim21_time}
\end{figure}

Figure~\ref{fig:sim21_time} supports these conclusions by providing the overall computation times (including aggregation and optimisation) for each model in
Figure~\ref{fig:sim2_pred}. 
We observe that the mean computational time required for the univariate symbolic model is significantly lower than for the multivariate model with full data, with an increasing disparity as the sample size $N$ increases, as the number of terms in the histogram-based OvR model depends on the histogram construction and not $N$ (and so is approximately constant in these plots). In addition to its prediction superiority, $L_{\tSO}^{(1)}$ also computationally outperforms the subsampling approach of \citet{wang2018}.
Note that Figure~\ref{fig:sim21_time} indicates that $L_{\tSO}^{(2)}$ (dotted black line) is computationally more demanding than using the full data (solid black line), making it superfluous in this setting. However note that as computation for $L_{\tSO}^{(2)}$ is constant in $N$, there will be some value $N_0$ such that if $N>N_0$ then the computational overheads for $L_{\tSO}^{(2)}$ will be more efficient than for the full data analysis.

\begin{figure}[t!]
\centering  
\includegraphics[page=1,width=0.4\textwidth]{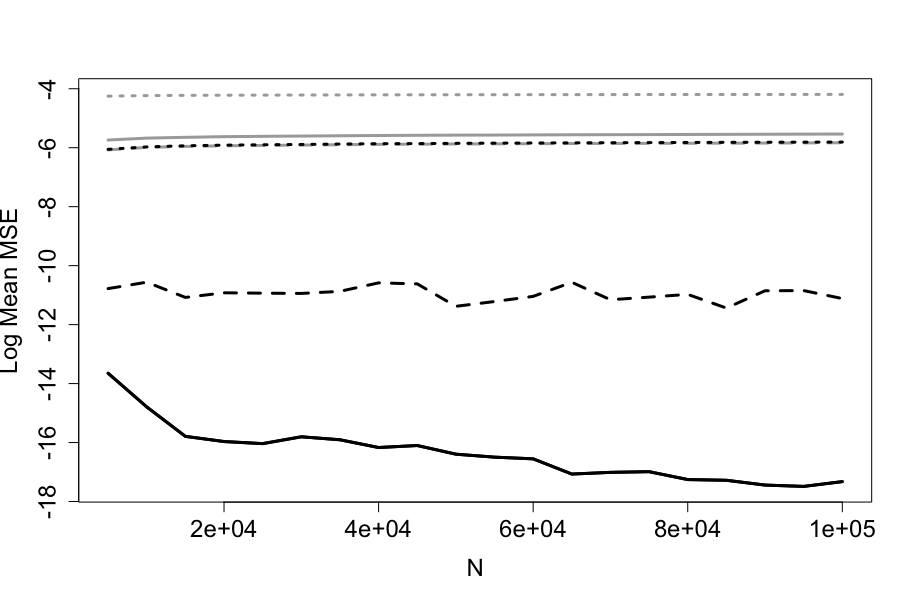}
\includegraphics[page=1,width=0.4\textwidth]{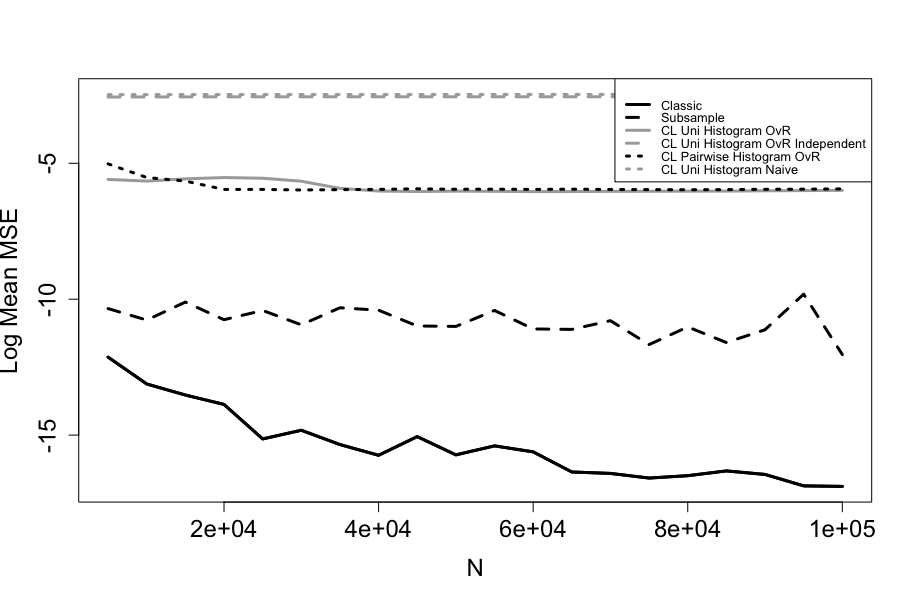}
 \caption{MMSE for the multinomial model on the full data (solid black line), subsampled data (dashed black line) and the histogram-based OvR model using $L_{\tSO}^{(1)}$ with independence assumption (dashed grey line), $L_{\tSO}^{(1)}$ with correlations (solid grey line), $L_{\tSO}^{(2)}$ (dotted black line) and the naive composite likelihood model (dotted grey line) as a function of the number of datapoints $N$. The covariates are generated from $8$-dimensional skew-normal distributions, considering zero (left) and non-zero (right) correlation parameters. Results are based on $1\,000$ replicate analyses.
  }
\label{fig:sim21_MSE}
\end{figure}

Figure~\ref{fig:sim21_MSE}  explores parameter estimator performance via
the mean mean squared error (MMSE) of a model's MLE, $\hat{\theta}^\textrm{Model}$, 
defined as $\textrm{MMSE}(\hat{\theta}^\Model) = S^{-1} \sum_{s=1}^S\| \hat{\theta}^\Model_s -\theta^{\textrm{true}}_s\|^2$, where $\|\cdot\|$ denotes the Euclidean norm, $\theta^{\textrm{true}}_s$ is the true parameter vector, and $S=1\,000$ the number of replicate analyses. Figure~\ref{fig:sim21_MSE} 
demonstrates that  subsampling methods perform better than the histogram-based methods if a low MMSE is desired. Using $L_{\tSO}^{(1)}$ instead of the naive composite likelihood analysis leads to a lower MMSE, with the results further improving if the covariate correlations are included.

\begin{figure}[t!]
\centering
\includegraphics[page=1,width=0.46\textwidth]{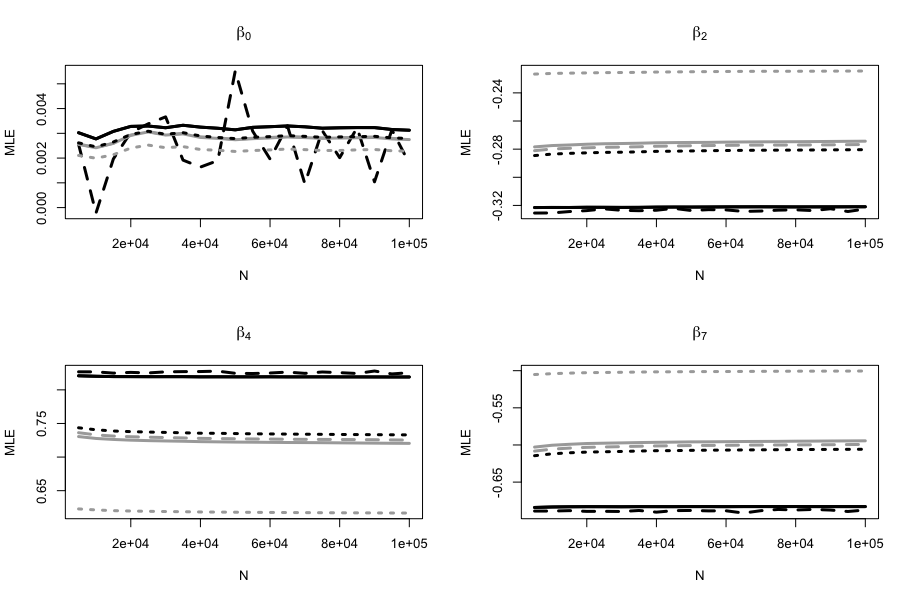}
\includegraphics[page=1,width=0.46\textwidth]{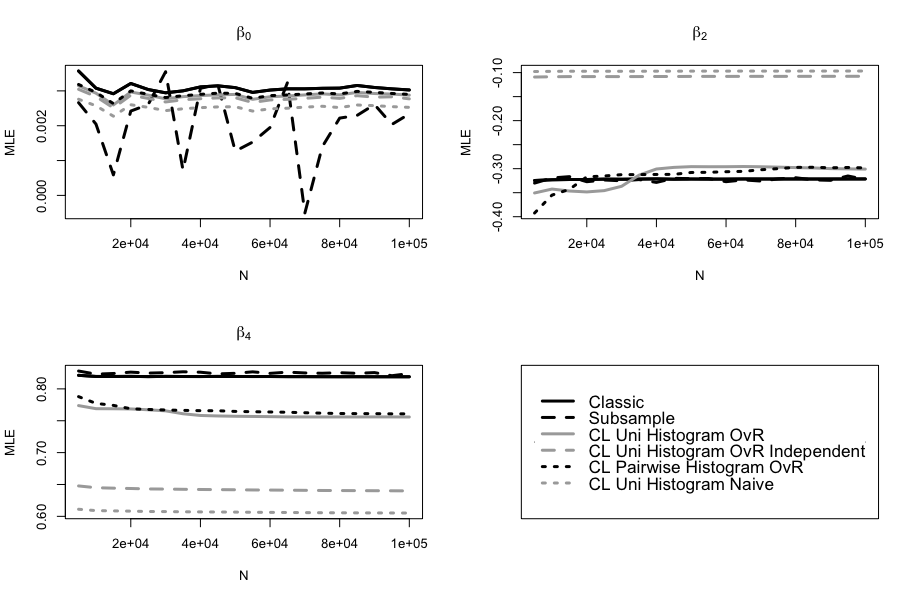}
 \caption{Mean MLEs using the multinomlial model on the full data (solid black line), subsampled data (dashed black line) and the histogram-based OvR model using $L_{\tSO}^{(1)}$ with independence assumption (dashed grey line), $L_{\tSO}^{(1)}$ with correlations (solid grey line), $L_{\tSO}^{(2)}$ (dotted black line) and the naive composite likelihood model (dotted grey line) as a function of the number of replicates $N$. The covariates are generated from $8$-dimensional skew-normal distributions, considering zero (left two columns) and non-zero (right two columns) correlation parameters. Results are based on $1\,000$ replicate analyses.
 }
\label{fig:sim2_param}
\end{figure}

Figure \ref{fig:sim2_param} explores parameter estimate accuracy further, displaying the mean estimates for a selection of the regression parameter over the $1\,000$ replicate analyses. The estimates obtained from $L_{\tSO}^{(1)}$ are much closer to those of the full model analysis than that of the naive composite likelihood analysis, with accuracy improving if covariate correlations are incorporated into the model. While the subsampling method provides more accurate parameter estimates, the better performance of the histogram-based models for predictions can potentially be explained by the fact that the histogram-based models incorporate the entire dataset, whereas a subsampling scheme can still potentially omit important observations. For predictions and the binary model, an observation $x_n$ is assigned to class $1$ if $\beta^\top x_n<0$, and class $2$ otherwise. Consequently, the same predictions are obtained from any parameter vector $m\beta$, $m>0$ for any dataset. 
In this case, the histogram-based models are more accurately estimating the model parameters to proportionality compared to the subsampling scheme, despite having a larger MSE.

In summary, this experiment suggests that if predictions are desired for logistic regression models, histogram-based solutions can be more accurate and computationally more efficient than subsampling-based methods, such as that in \citet{wang2018}. The use of bivariate histograms to represent the covariate information improves the prediction of the response outcomes, but at an often impractical computational cost compared to univariate histograms. This simulation study suggests that marginally aggregating covariates into univariate histograms, in combination with knowledge of covariate correlations, provides the best trade off between accuracy and speed. 

Methods based on e.g.~\cite{beranger+ls18} can be used to estimate correlations from multidimensional histograms ($j \geq 2$). However only variance estimation (not correlations) can be obtained from univariate marginal histograms, meaning that correlations need to be obtained prior to data aggregation. Given the small computational 
costs of estimating covariance matrices, even for large $N$ data sets, this does not affect the efficiency of the proposed method.
However, if an estimate of the observed covariance matrix is unavailable, this would force the analyst to assume independence between covariates, leading to a decrease in performance (if large correlations are present).


\section{Real data analyses}
\label{realdata}

We illustrate the applicability of our proposed methodology to two real data problems.
We first consider a logistic regression problem where the goal is to distinguish between a process where new supersymmetric particles are produced and a background process. Secondly we tackle a multinomial regression problem which consists of predicting crop types based on satellite-based pixel observations.


\subsection{Supersymmetric benchmark dataset}
\label{sec:HEP}

The Supersymmetry dataset (SUSY) is  available from the Machine Learning Repository \citep{dua2019} and comprises  $5$ million Monte Carlo observations generated by \cite{super2014}. The binary response variable ($K=2$) discriminates a signal process which produces supersymmetric particles from a background process which does not. There are 18 features ($D=18$); the first 8 are low-level features representing the kinematic properties measured by the particle detectors, while the remaining 10 are high-level features derived as function of the previous 8 by physicists to help discriminate between the two outcomes. This dataset was analysed by \cite{wang2018} to test their optimal subsampling scheme for logistic regression. Following \cite{wang2018}, we consider a training dataset of $4\,500\,000$ randomly chosen observations, and a test dataset with the remaining $500\,000$ observations.    

Following the conclusions from Section \ref{sec:sim}, we fit the histogram-based OvR model using univariate marginal histogram aggregates ($L_{\tSO}^{(1)}$). While until now the focus has been on histograms with random counts (fixed bins), here we fit $L_{\tOO}^{(1)}$ to explore the  performance of using random bin (fixed counts) histograms. For an integer $B$, we construct histograms for each covariate by partitioning the data into $B$ bins with 
roughly equal counts,  while avoiding the highest and lowest 1\% of covariate values to reduce bin sensitivity to outliers. E.g.~for $B=4$ we would use the  $0.01, 0.25, 0.5, $ and $0.99$ empirical quantiles to construct the histogram for each covariate,   and create additional bins on the end points of the range of the data to accommodate the outliers.

The likelihood functions are optimised using Lasso regularisation with $10$-fold cross-validation and, for $L_\tOO^{(1)}$,  for a range of values of $B$. Prediction accuracies obtained on the test dataset and the optimisation times (in seconds) on the training dataset are reported in Table~\ref{tab:SUSY}. For the histogram-based models, there is an increase in prediction accuracy as the number of bins increases, which is as expected since these are more informative summaries. The improvement in prediction accuracy slows down at around the $B=12$ bin mark, whereas  the computation time naturally increases with the number of bins. When implementing the subsampling methodology of  \cite{wang2018} 
we obtain a prediction accuracy of $78.2\%$ with a computation time of $70.2$ seconds, i.e.~about 3--4 times more computation than for the histogram-based models with $B=25$. 
That is, the histogram-based methods offer as good prediction accuracy with much smaller computational overheads compared to state-of-the-art subsampling approaches.
\begin{table}
\small
\begin{tabular}{ cc@{\hskip 0.1in}c@{\hskip 0.1in}c@{\hskip 0.1in}c@{\hskip 0.1in}c@{\hskip 0.1in}c@{\hskip 0.1in}c@{\hskip 0.1in}c@{\hskip 0.1in}c@{\hskip 0.1in}c@{\hskip 0.1in}c@{\hskip 0.1in}c@{\hskip 0.1in}c@{\hskip 0.1in}c@{\hskip 0.1in}c} 
& & \multicolumn{13}{c}{Bins} \\
 \cline{3-15}
Likelihood  & & 6 & & 8 & & 10 & & 12 & & 15 & & 20 & & 25 \\
\cline{1-1} \cline{3-3} \cline{5-5}  \cline{7-7} \cline{9-9} \cline{11-11} \cline{13-13} \cline{15-15}
$L_{\tOO}^{(1)}$ & &  75.9& & 77.8& & 77.9& & 78.0& & 78.0& &78.1 & & 78.1 \\
 & &  (19.2) & & (15.3) & & (18.9) & & (19.4) & & (20.5) & & (20.1) & & (21.6) \\
$L_{\tSO}^{(1)}$  & &  76.2 & & 77.4& & 78.0& & 78.2& & 78.2 & & 78.2 & & 78.3 \\ 
  & &  (10.3) & & (10.5) & & (12.0) & & (12.9) & & (14.1) & & (15.2) & & (15.8) \\ 
  Subsampling&&&&&&&&&&&&&&& 78.2\\
   \cite{wang2018}&&&&&&&&&&&&&&& (70.2)\\
 \hline
\end{tabular}
\caption{ Percentage prediction accuracy with computing time (in seconds) for the Supersymmetry dataset, using histograms with $B$ bins per margins, compared to the subsampling approach of \cite{wang2018}.}
\label{tab:SUSY}
\end{table}

\subsection{Crop type dataset}
\label{sec:crop}

\begin{figure}
\centering
\includegraphics[page=1,width=0.45\textwidth]{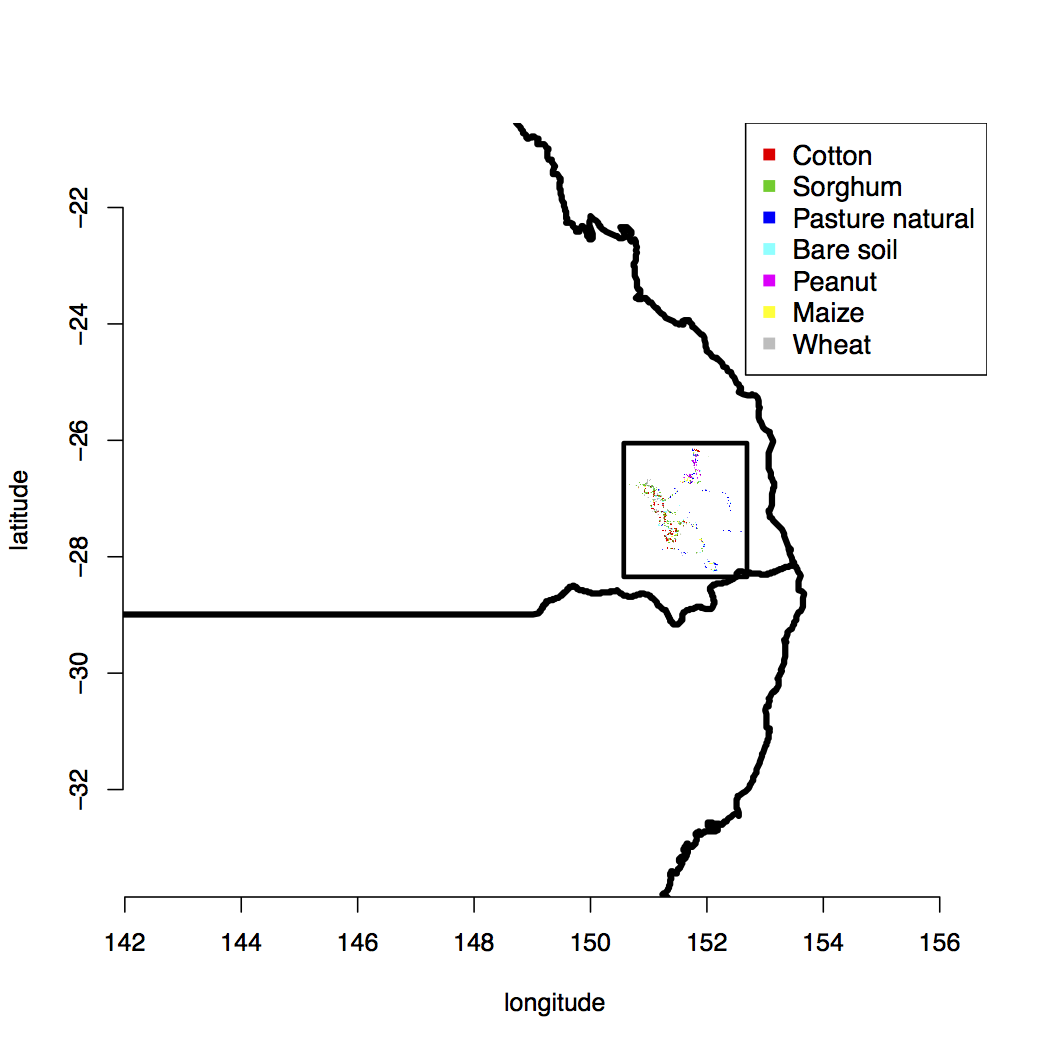}
\includegraphics[page=1,width=0.45\textwidth]{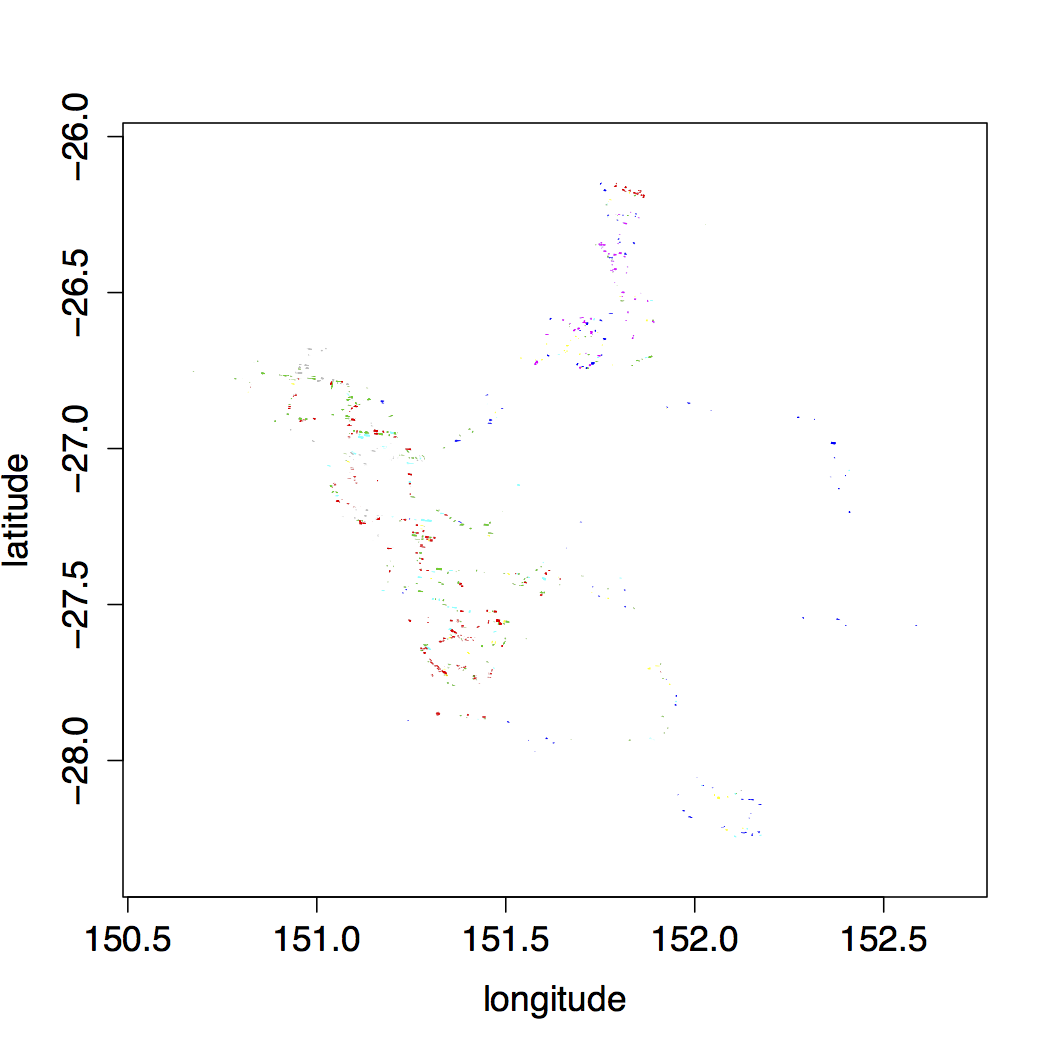}
 \caption{\small The crop type dataset with different colours for each crop. Left: Location of the study area in the state of Queensland on the east coast of Australia. Right: farm location and crop type detail.
 }
\label{fig:real_map}
\end{figure}

We examine a crop type dataset \citep{QUT} which consists of $247\,210$ observations, each representing a $25\times25$m$^2$ pixel located over farmland across the state of Queensland, on the east coast of Australia (Figure \ref{fig:real_map}). For each pixel the ground-truth crop type is available (observed at one of three possible times) as well as numerous vegetation indices, based on reflectance data taken from a LANDSAT 7 satellite. The aim of this analysis is to predict the crop type based on the vegetation indices. After selecting the most meaningful covariates 
by iteratively removing variables with correlations greater than $0.85$, we retained $D=7$ variables corresponding to various colour reflectances measured by the satellite and functions of these indicators.

As poor prediction accuracy of classes with low numbers of observations is a well known issue,
we only retain crop types that are observed more than $10\,000$ times, reducing the dataset to $234\,485$ observations. The set of possible outcomes of our multinomial response variable $Y = ``\textrm{Crop type}"$ is then $\Omega =\{ \textrm{Bare Soil}, \textrm{Cotton}, \textrm{Maize}, \textrm{Pasture Natural}, \textrm{Peanut}, \textrm{Sorghum}, \textrm{Wheat} \}$ and thus $K=7$. 
The resulting dataset is identical to the one used in a previous analysis in \cite{QUT} which used the standard multinomial model $L_\tM(\bx,\by;\bbeta)$.

As the approximate composite likelihood relies on the assumption of a linear relationship between the predictor variables, we use the R package $\tt bestNormalize$  to select the best transformation to achieve approximate predictor Gaussianity, according to the Pearson P-test statistic.
The dataset is randomly partitioned into a training dataset of size $200\,000$ used for parameter estimation and a test dataset with the remaining $34\,485$ observations to evaluate the prediction accuracy. We perform constrained likelihood optimisation with a Lasso regularisation, and use $10$-fold cross validation to determine the best regularisation parameter.

\begin{table}
\begin{tabular}{ c@{\hskip 0.1in}c@{\hskip 0.1in}c@{\hskip 0.1in}c@{\hskip 0.1in}c@{\hskip 0.1in}c@{\hskip 0.1in}c@{\hskip 0.1in}c@{\hskip 0.1in}c@{\hskip 0.1in}c@{\hskip 0.1in}c@{\hskip 0.1in}c@{\hskip 0.1in}c@{\hskip 0.1in}c@{\hskip 0.1in}c@{\hskip 0.1in}c@{\hskip 0.1in}c@{\hskip 0.1in}c@{\hskip 0.1in}c}
& & & & \multicolumn{12}{c}{Bins}& \\
\cline{5-15}

Crop type & & $N_k$ & & $6$ & & $8$ & & $10$ & & $12$ & & $15$ & & $20$ & &    $L_\tM(\bx,y;\bbeta)$ \\
\cline{1-1} \cline{3-3} \cline{5-5} \cline{7-7} \cline{9-9} \cline{11-11} \cline{13-13} \cline{15-15} \cline{17-17}
Cotton                & & $\phantom{0}72\,450$ & & 90.9 & & 93.6 & & 92.4 & & 91.4 & & 91.3 & & 91.3 &  &  92.6\\ 
Sorghum            & & $\phantom{0}66\,751$ & & 78.8& & 76.2  & & 78.6 & & 79.0 & & 78.4 & & 77.7 &  &  80.3 \\ 
Pasture Natural & & $\phantom{0}27\,479$ & &76.1& & 81.2 & & 81.4 & & 81.3 & & 81.8 & & 82.2 &  &  77.6 \\ 
Bare Soil            & & $\phantom{0}26\,173$ & & 89.5 & & 90.2 & & 90.4 & & 90.0 & & 89.8& & 90.1 &  &  91.0 \\ 
Peanut               & & $\phantom{0}17\,868$ & & 80.2 & & 79.7 & & 79.9 & & 79.7 & & 80.2& & 79.8 &  & 82.9 \\ 
Maize 		& & $\phantom{0}12\,986$ & & \phantom{0}0.6 & & \phantom{0}1.3& &  \phantom{0}5.0& & 11.9 & &  \phantom{0}5.5 & &  \phantom{0}7.3 &   & 14.2 \\ 
Wheat 		& & $\phantom{0}10\,778$ & & \phantom{0}6.7 & & 10.4& & \phantom{0}9.9 & & \phantom{0}8.7 & & \phantom{0}9.9 & & 10.4 &  &  10.3 \\
\cline{1-1} \cline{3-3} \cline{5-5} \cline{7-7} \cline{9-9} \cline{11-11} \cline{13-13} \cline{15-15} \cline{17-17} 
Overall & & $234\,485$ & & 75.8 & & 76.8 & & 77.3 & & $77.2$ & & $77.3$ & & $77.4$ & & 78.1 \\
 & &  & & (49) & & (70) & & (73) & & (94) & & (97) & & (117) &  & (1263) \\
\hline 

\end{tabular} 
\caption{ Crop specific and overall prediction accuracies (\%) using univariate marginal histograms with $B$ bins. The likelihood optimisation times (in seconds) are reported in the last row. The full model is the standard multinomial likelihood $L_\tM(\bx,\by;\bbeta)$ \eqref{eq:MLR} with Lasso regularisation, as implemented by \cite{QUT}.}
\label{tab:real1} 
\end{table}

Table~\ref{tab:real1} presents the prediction accuracies for the $L_{SO}^{(1)}$ model for each crop type and the overall prediction accuracy when the covariate information is collapsed into univariate marginal histograms with $B=6,8,10,12,15$ and $20$ bins.
The last column of Table~\ref{tab:real1} provides a comparison with the full data multinomial likelihood ($L_\tM(\bx,\by;\bbeta)$) using the same Lasso regularisation, as implemented in the original analysis by  \cite{QUT}.
The overall  and crop-specific prediction accuracies have achieved good predictive performance compared to the full data multinomial model analysis using only 
$B\approx 10$--$12$ bins. 
Two particular crops produce notable results. The prediction accuracies for Maize are around 7.3\%  for the histogram-based analysis compared to the $\sim$14.2\% accuracy of the full-data analysis. While both of these are low due to this crop being the least well represented of all crops in the study (less than 6\% of all observations), and perhaps lowly informative vegetation indices for this crop, the $\sim$50\% predictive underperformance for the histogram-based analysis suggests that categories with less data in a model are more sensitive to the degree of binning than those categories with larger representation in the dataset (although this is less apparent for Wheat). 
 
In the case of Pasture Natural, the histogram-based prediction accuracies are even higher (at $\sim$82\%) than for the full data analysis (77.6\%). While this is not immediately understandable intuitively, in that by constructing histograms information in the dataset is certainly being lost and so performance should perhaps always be worse, the difference is only $\sim$4\%, and moreover the likelihoods are not directly comparable in the sense that the limit of the approximate composite likelihood $L_\tSO^{(j)}(\bs;\bbeta)$ as $B\rightarrow\infty$ is not the full data multinomial model $L_\tM(\bx,y;\bbeta)$ as used in \cite{QUT}. So here the discrepancy is that the two likelihoods simply have different performances for these data. This argument notwithstanding, proponents of symbolic data analysis sometimes ascribe to the idea that inference using the `shape' of the data (that is, the histogram summary) may sometimes be more useful in an analysis than the full underlying dataset (Edwin Diday, personal communication).

Overall, while the histogram-based analysis gives comparable prediction accuracies to the full data analysis, the real gains are in the computational overheads required for each model. The full multinomial analysis takes considerably longer  (more than 13$\times$ the $B=12$ analysis) to implement than the histogram based analysis. 
Finally, note that while the computational savings here are substantial, this dataset only contains $N=234\,485$ observations. For larger datasets the computational overheads will skyrocket for the standard multinomial model analysis (where computation is proportional to $N$), and yet will remain roughly constant for the histogram-based approach (where computation is proportional to the number of bins).

\section{Discussion}

In this article, we have developed a novel approach for classifying binary and multinomial random variables 
that alleviates the computational bottleneck that arises with very large  datasets. The strategy 
relies on concepts from the field of symbolic data analysis
\citep{beranger+ls18},
aggregating the covariate data into histogram-valued random variables which have lower computational overheads to analyse and store, albeit with some loss of information.
When computation for any histogram bin is larger than that for the standard likelihood contribution of the datapoints within that bin, the standard likelihood contribution for these datapoints can be used instead.
However, because high-dimensional histograms are not efficient distributional summaries,
we additionally introduced an approximate composite likelihood  methodology, which quantitatively builds on the qualitative results of \cite{cramer2007}. The individual components of the approximate composite likelihood are constructed from marginal histograms derived from the full $D$-dimensional histogram.
This concept of approximate composite likelihoods for logistic regression does not solely apply to aggregated data and can be used in more general settings.

We have demonstrated through simulation studies and real data analyses that these histogram-based strategies can produce fitted models that have comparable prediction accuracies to the standard full data analysis, but at a much lower computational cost, even compared to state-of-the-art computational techniques for logistic regression such as subsampling \citep{wang2018}. On the down side, the resulting parameter estimates are biased, though not as much as for naive composite likelihood-based approaches.
 
One aspect of implementing histogram-based inference that we have not explored is identifying a principled method of constructing histograms for subsequent analysis.
If the number of bins is too low, then important information in the data will be lost and model predictions may be poor (see e.g.~Figure \ref{fig:sim1_pred}). In contrast, as the number of bins becomes large, the inferential accuracy can approach the level of the full data analysis (within the context of the inferential model being used). However, the price of more accurate inference is an increase in the computational costs. A simple diagnostic to choose the number of bins approach is apparent from the real data analyses in Section \ref{realdata}.
Here the number of bins could be increased until some quantity of interest -- such as the model prediction accuracy, or the MLE, or it's standard error -- does not change drastically when increasing the number of bins further. Of course, this procedure is very ad-hoc, and so developing a more theoretically justified and principled approach would be highly beneficial.
Such an approach could therefore consider balancing computational complexity and inferential accuracy, or alternatively by minimising a loss function constructed over some useful criterion. 

Similarly, the assumption of equal bin widths considered here could be relaxed by developing selection criteria for bin locations.  
As suggested by the Associate Editor, regression trees over covariate space offer one possibility to efficiently represent the data by identifying easily integrable (axis-parallel) regions over covariate space.
One advantage of using equally-spaced bins, however, is that bin locations are computationally cheap to obtain, requiring just the minimum and maximum values for each variable. Conversely, any optimal bin-choice method will be more computationally expensive to implement, and so any gains in accuracy must be balanced with the increase in computation.
This is clearly an important component of the current methodology, and is the focus of current research.

%
\section*{Acknowledgements}
This research is supported by the Australian Research Council through the Australian Centre of Excellence for Mathematical and Statistical Frontiers  (ACEMS; CE140100049), and the Discovery Project scheme (FT170100079).
\appendix
 
 \section*{Code and Data}
 
 The code and dataset used in this paper can be found at:\\
{\tiny
\begin{tt} 
 https://media.githubusercontent.com/media/borisberanger/academic-kickstart/master/static/zip/WBS\_2020.zip
 \end{tt}
 }
 
\section{Appendices}

\subsection{Proof of Proposition \ref{prop1}}
\label{proofprop1}
We utilise the arguments presented in \cite{beranger+ls18} to derive of Proposition \ref{prop1}. Note that if the underlying microdata $\bold X=( X_1,..., X_N)$ are i.i.d.~then the $K$ subsets $\bold X^{(1)},\ldots,\bold X^{(K)}$ are similarly i.i.d. For histogram-valued data,
\begin{align*}
f(\bold s|\bx, \by,\vartheta) &= \prod_{k=1}^Kf( s_k|\bx^{(k)},\vartheta),
\end{align*}
where 
$$f( s_k|\bx^{(k)},\vartheta) =\prod_{\bb_k=\bold 1_k}^{\bB_k}\boldsymbol{1} \left\{\sum_{n=1}^{N_k}\boldsymbol{1}\{x_n^{(k)}\in\bUpsilon_{\bb_k}\} = s_{\bb_k}  \right\}.$$
 For the symbolic, histogram-based model, we therefore obtain
\begin{align*}
L_{SM}(\bold s;\beta) &= \int_{\dom_{\bold X}}L_{M}(\bx, \by;\beta)f(\bold s|\bx, \by,\vartheta)\der\bx\\
&=\prod_{k\in\Omega} \int_{\dom_{\bold X^{(k)}}}L_{M}(\bx^{(k)}, \by;\beta)f( s_k|\bx^{(k)},\vartheta)\der\bx^{(k)}\\
&= \prod_{k\in\Omega}\prod_{n=1}^{N_k}\prod_{\bb_k=\bold 1_k}^{\bB_k}\left (\int_{\dom_{X_{n}^{(k)}}}L_{M}(x_{n}^{(k)}, y_n;\beta)\der x_{n}^{(k)}\right )^{\boldsymbol{1}\{x_{n}^{(k)}\in\bUpsilon_{\bb_k}\}}\\
&\propto \prod_{k\in\Omega}\prod_{\bb_k=\bold 1_k}^{\bB_k}\left(\int_{\bUpsilon_{\bb_k}}P_{M}(y=k| X=x)\der x\right )^{ s_{\bb_k}}.
\end{align*}
Similarly, for the histogram-based OvR model, we obtain
\begin{align*}
L_{SO}(\bold s;\beta) &= \int_{\dom_{\bold X}}L_{O}(\bx, \by;\beta)f(\bold s|\bx,\by,\vartheta)\der\bx\\
&= \prod_{k\in\Omega} \int_{\dom_{\bold X^{(k)}}}L_{O}(\bx^{(k)}, \by;\beta)f( s_k|\bx^{(k)},\vartheta)\der\bx^{(k)}\\
&= \prod_{k\in\Omega} \prod_{n=1}^{N_k}\prod_{\bb_k=\bold 1_k}^{\bB_k}\left (\int_{\dom_{X_{n}^{(k)}}}L_{O}(x_{n}^{(k)}, y_n;\beta)\der x_{n}^{(k)}\right )^{\boldsymbol{1}\{x_{n}^{(k)}\in\bUpsilon_{\bb_k}\}}\\
&\propto \prod_{k \in \Omega} \prod_{\bb_k= \bOne_k}^{\bB_k} 
\left(
\int_{\bold \Upsilon_{\bb_k}}P_{\tO}(Y=k| X=x) \der x
\prod_{k' \in \Omega\backslash \{k\} } 
\int_{\bold \Upsilon_{\bb_k}}P_{\tO}(Y\neq k'| X=x) \der x
\right)^{s_{\bb_k} }.
\end{align*}

\subsection{Proof of Proposition \ref{prop:NewProp}}
\label{SeparationProp}
We now show that if there is neither complete nor quasi-complete separation of the set of histograms $\bs$, then $L_{SO}(\bold s;\beta)$ and $L_{SM}(\bold s;\beta)$ have unique global maxima. The following arguments are analogous to those proposed by \cite{Adelin1984}.
Suppose that there is complete separation exhibited by the histogram dataset for the $k^{th}$ class, according to Definition \ref{def4}.

As a result, the complete separation property holds for all vectors $\beta_k=a_kb_k$ for $a_k>0$. We now examine the behaviours of the integrals of the $P_{\tO}(Y=k | X)$ and $P_{\tO}(Y\neq k | X)$ terms in the likelihood functions \eqref{eq:SM} and \eqref{eq:SO}. Using the mean value theorem, and given $a_kb_k^\top  x^*_{\bb_j}>0$ if $j=k$ and $a_kb_k^\top  x^*_{\bb_j}<0$ if $j\neq k$ for all non-empty bins and $a_k>0$, we obtain
\begin{align*}
\lim_{a_k\rightarrow \infty}\int_{\bold \Upsilon_{\bb_k}}P_{\tO}(Y=k | x)\der x &= \lim_{a_k\rightarrow \infty}\int_{\bold \Upsilon_{\bb_k}}\frac{e^{a_kb_k^\top  x}}{1+e^{a_kb_k^\top  x}}\der x\propto \lim_{a_k\rightarrow \infty}\frac{e^{a_kb_k^\top  x^*_{\bb_k}}}{1+e^{a_kb_k^\top  x^*_{\bb_k}}}=1\\
\lim_{a_k\rightarrow \infty}\int_{\bold \Upsilon_{\bb_k}}P_{\tO}(Y\neq k | x)\der x &= \lim_{a_k\rightarrow \infty}\int_{\bold \Upsilon_{\bb_k}}\frac{1}{1+e^{a_kb_k^\top  x}}\der x \propto \lim_{a_k\rightarrow \infty}\frac{1}{1+e^{a_kb_k^\top  x^*_{\bb_k}}}=1,
\end{align*}
where $x^*_{\bb_k}$ is some point inside $\bUpsilon_{\bb_k}$. Each integral therefore approaches a constant for all bins as $a_k$ increases, meaning the maximum value of each likelihood function is attained at the boundary of the parameter space, i.e.~$\hat{\beta}_k=\infty$.

Now suppose there is quasi-complete separation exhibited by the histogram dataset, 
according to Definition \ref{def4}.
Continuing with the previous notation, denote $A_k^{D+1}$ as the set of all vectors $b_k$ that satisfy the complete separation condition, meaning that $A_k^{D+1}$ is a convex set. Denote the parameter vector $\alpha_k(a_k) = c_k+a_kb_k$, where $a_k>0$ and $a_k\in A^{D+1}$. Consequently,
$$P_{\tO}(Y=k | x)= \frac{e^{\alpha_k(a)^\top  x}}{1+e^{\alpha_k(a_k)^\top  x}}.$$
The log-likelihood for the component of the OvR model \eqref{eq:SO} estimating the parameters for the $k^{th}$ class, $\beta_k$, can therefore be expressed as 
\begin{align*}
\log L_{SO}^k(\bold s;\beta_k) &=  \sum_{\bb_k=\bOne_k}^{\bB_k}s_{\bb_k}\log \int_{\bold \Upsilon_{\bb_k}}P_{\tO}(Y=k | x)\der x+\sum_{k'\in \Omega \backslash \{k\}}\sum_{\bb_{k'}=\bOne_{k'}}^{\bB_{k'}}s_{\bb_{k'}}\log \int_{\bold \Upsilon_{\bb_{k'}}}P_{\tO}(Y\neq k | x)\der x\\
&=\sum_{\bb_k=\bOne_k}^{\bB_k}s_{\bb_k}\log \int_{\bold \Upsilon_{\bb_k}} \frac{e^{c_k^\top  x+a_kb_k^\top  x}}{1+e^{c_k^\top  x+a_kb_k^\top  x}} \der x+\sum_{k'\in \Omega \backslash \{k\}}\sum_{\bb_{k'}=\bOne_{k'}}^{\bB_{k'}}s_{\bb_{k'}}\log \int_{\bold \Upsilon_{\bb_{k'}}}\frac{1}{1+e^{ c_k^\top  x+a_kb_k^\top  x}}\der x.
\end{align*}
Given that $b_k^\top x > 0$ for all $x\in \bUpsilon_{\bb_k}$ such that $s_{\bb_k}>0$, the function $\frac{e^{c_k^\top  x+a_kb_k^\top  x}}{1+e^{c_k^\top  x+a_kb_k^\top  x}}$ is monotonically increasing with $a_k$  for all $x\in \bUpsilon_{\bb_k}$ such that $s_{\bb_k}>0$. Consequently, 
$$\sum_{\bb_k=\bOne_k}^{\bB_k}s_{\bb_k}\log \int_{\bold \Upsilon_{\bb_k}} \frac{e^{c_k^\top  x+a_kb_k^\top  x}}{1+e^{c_k^\top  x+a_kb_k^\top  x}} \der x$$
is monotonically increasing with increasing $a_k$. Similarly, given that $b_k^\top x < 0$ for all $x\in \bUpsilon_{\bb_{k'}}$ such that $s_{\bb_{k'}}>0$ and $k'\neq k$
$$\sum_{k'\in \Omega \backslash \{k\}}\sum_{\bb_{k'}=\bOne_{k'}}^{\bB_{k'}}s_{\bb_{k'}}\log \int_{\bold \Upsilon_{\bb_{k'}}}\frac{1}{1+e^{ c_k^\top  x+a_kb_k^\top  x}}\der x$$
is monotonically increasing with increasing $a_k$. Therefore the log-likelihood function 
$\log L_{SO}^k(\bold s;\beta)$ 
is monotonically increasing with $a_k$, and the maximum value is attained at the boundary of the parameter domain, i.e.~$\hat\beta_k=\infty$. 

The above arguments show that if there is complete or quasi-complete separation for any class $k\in\Omega$, then there is no unique MLE for the symbolic OvR model. Using similar arguments, it can be shown that the maximum value for the symbolic multinomial likelihood 
$\log L_{SM}(\bold s;\bbeta)$ 
is attained at $\hat\bbeta=(\infty,\ldots,\infty)^\top$ if there is either complete or quasi-complete separation in the data. 

By the mean value theorem, 
\begin{align*}
L_{SM}(\bold s;\bbeta) &\propto \prod_{k\in\Omega}\prod_{\bb_k=\bold 1_k}^{\bB_k}\left(\int_{\bUpsilon_{\bb_k}}P_{M}(y=k| X=x)\der x\right )^{ s_{\bb_k}}\\
&\propto \prod_{k\in\Omega}\prod_{\bb_k=\bold 1_k}^{\bB_k}P_{M}(y=k| X=x^*_{\bb_k})^{ s_{\bb_k}},
\end{align*}
where $x^*_{\bb_k}\in \bUpsilon_{\bb_k}$ is some point located inside the $\bb_k^{th}$ bin. The symbolic multinomial likelihood $L_{SM}(\bold s;\bbeta)$ is therefore proportional to the classical likelihood for some dataset $\bx^*$, consisting of $\sum_{k\in\Omega}\sum_{\bb_k=\bold 1_k}^{\bB_k}\boldsymbol{1} \{s_{\bb_k}>0\}$ distinct values $x^*_{\bb_k}$, each appearing $s_{\bb_k}$ times, $\bb_k=\bOne_k,\ldots,\bB_k$, $k\in\Omega$. \cite{Adelin1984} proved that $L_\tM (\bx, \by; \bbeta) $ is a closed convex function. Therefore, $L_\tM (\bx^*, \by; \bbeta) $ and subsequently $L_{SM}(\bold s;\bbeta)$ are closed convex functions. Similar arguments can be used to show the closed convex nature of $L_{SO}(\bold s;\bbeta)$. As a result, if the histogram-valued data does not exhibit complete separation or quasi-complete separation (Definition \ref{def4}), then there is a unique global maximum of $L_{SM}(\bold s;\bbeta) $. Similarly, if the histogram-valued data does not exhibit complete separation or quasi-complete separation for any class $k\in\Omega$ (Definition \ref{def3}), then there is a unique global maximum of $L_{SO}(\bold s;\bbeta)$.

\subsection{Proof of Proposition \ref{prop:composite}}
\label{proofprop:composite}
We first use the arguments in \cite{cramer2007,Wooldridge2002} to derive $L_O^{(j)}$ and $L_{SO}^{(j)}$ using the latent variable formulation of the logistic regression model (equivalent to the log odds formulation in Section \ref{sec:log_reg}). Consider a binary $(K=2)$ logistic regression model, such that the OvR and multinomial model are equivalent. The latent variable formulation for the logistic regression model based on the $\bi^{th}$ subset of variables can be written as
\begin{equation}
Y_{n}^{*} =  \bold \beta^{\bi \top} X_n^{\bi}+e_{n}^{\bi},\qquad n=1,\ldots,N,
\label{eq:lat}
\end{equation}
where $Y_{n}^{*}$ is an unseen latent variable and $e_{n}^{\bi}$ is an error term following a logistic distribution. Classification is then achieved by setting $Y_n=1$ if $Y_{n}^*<0$ and $Y_n =2$ otherwise. In the full model
$$Y_{n}^* =  \bold \beta^{\top} X_n^{}+u_{n},$$
where $u_{n}$ follows a logistic distribution with zero mean and unit variance. In the smaller model (\ref{eq:lat}) indexed by $\bi$, the omitted terms are absorbed into the error term. That is
$$e_{n}^{\bi} = u_{n}+\sum_{i'\in \mathcal I_1^{-\bi}}\beta_{}^{i'}X_{ni'}.$$
Suppose that there are correlations between included and omitted variables for the model based on subset $\bi$, and we can express each omitted variable as a linear function of the included variables, i.e.~$X^{i'} = \alpha_{\bi i'}^\top X^{\bi} + \epsilon_{\bi i'}$, as described in Section \ref{sec:comp_improv}. W.l.o.g.~we can assume the $i^{th}$ variable has zero mean and variance given by $\sigma_i^2$, and that the covariance between variables $i$ and $i'$ is given by $\sigma_{ii'}$. The error term $e_{n}^{\bi}$ can therefore be expressed as
$$e_{n}^{\bi} = u_{n}+\sum_{i'\in \mathcal I_1^{-\bi}}\beta_{}^{i'}\left ( \alpha_{\bi i'}^\top X_n^{\bi} + \epsilon_{n\bi i'}\right).$$
We now rewrite ($\ref{eq:lat}$) by absorbing the terms in $e_{n}^{\bi}$ that are dependent on the included variables into the model. That is
$$Y_{n}^{*} =  \sum_{i_1'\in \bi}\left (\beta_{}^{i_1'}+\sum_{i_2'\not \in \bi}\beta_{}^{i_2'}\alpha_{i_1'i_2'}\right )X_{ni_1'}+\tilde e_{n}^{\bi},$$
where 
$$\tilde e_{n}^{\bi} = u_{n}+\sum_{i' \not \in \bi}\beta_{}^{i'}\epsilon_{n\bi i'}.$$
\cite{Pingel2014} shows that if $X$ is distributed according to a logistic distribution with mean $0$ and variance $\frac{\pi^2}{3}$, then a standard normal distribution also fits the distribution of $X$ reasonably well. The similarities between the standard logistic density and a rescaled normal density have also been investigated by \cite{Jeffress1973}, \cite{Bowling2009} and \cite{Pingel2014}, whereby different values for the rescaling factor $C$ are proposed based on the criteria used to match the logistic and normal distributions. In practise any of these values can be used here, but we proceed with $\frac{\pi^2}{3}$ in the simulations and real data analyses in Sections \ref{sec:sim} and \ref{realdata} respectively. As a result, $\sum_{i' \not \in \bi}\beta_{}^{i'}\epsilon_{n\bi i'}$ is approximately normally distributed, and the error term $\tilde e_{n}^{\bi}$ is approximately logistically distributed with mean zero (i.e.~$\mathbb E(\tilde e_{n}^{\bi}) = 0$) and variance given by
$$\mbox{Var}(\tilde e_{n}^{\bi}) = 1 + \frac{\sum_{i_1' \in \mathcal{I}_1^{-\bi}} \left( \beta^{i_1'2} \lambda_{\bi i_1'}^2 + 2  \sum_{i_2' \in \mathcal{I}_1^{-\bi}, i_2' \neq i_1'} \beta^{i_1'} \beta^{i_2'} \lambda_{\bi i_1' i_2'} \right)}{\pi^2/3}.$$ In the full model,  
$$P(Y_n=2| X_n) = P(Y_{n}^*>0| X_n) = P(u_{n}<\beta_{}^{\top}X_{n})=\frac{\exp\{\beta_{}^{\top}X_{n}\}}{1+\exp\{\beta_{}^{\top}X_{n}\}},$$
and we obtain MLE's $\hat\beta$ for $\beta$ by maximising over the sum of this quantity over all observations. Note that $P(u_{n}<\beta_{}^{\top}X_{n})= P_{O}(Y_n=2| X_n)$, resulting in the equivalency between the latent and log odds formulations of the logistic regression model. For the omitted variable model, 
$$P(Y_n=2| X_n^{\bi}) = P(Y_{n}^{\bi*}>0| X_n^{\bi}) = P\left (\tilde e_{n}^{\bi}<\sum_{i_1'\in \bi}\left (\beta_{}^{i_1'}+\sum_{i_2'\not \in \bi}\beta_{}^{i_2'}\alpha_{i_1'i_2'}\right )X_{ni_1'}\right ).$$
Rescaling $\tilde e_{n}^{\bi}$ by its standard deviation gives us a random variable with approximately the same distribution as $u_{n}$, i.e.~$\tilde u_{n}=\frac{\tilde e_{n}^{\bi}}{\sqrt{\mbox{Var}(e_{n}^{\bi})}}$ will approximately follow a logistic distribution with zero mean and unit variance. As a consequence,
\begin{align*}
P(Y_n=2| X_n^{\bi}) &= P\left (\frac{\tilde e_{n}^{\bi}}{\sqrt{\mbox{Var}(\tilde e_{n}^{\bi})}}<\frac{\sum_{i_1'\in \bi}\left (\beta_{}^{i_1'}+\sum_{i_2'\not \in \bi}\beta_{}^{i_2'}\alpha_{i_1'i_2'}\right )X_{ni_1'}}{\sqrt{\mbox{Var}(\tilde e_{n}^{\bi})}}\right )\\
&\approx  P\left (\tilde u_{n}<\frac{\sum_{i_1'\in \bi}\left (\beta_{}^{i_1'}+\sum_{i_2'\not \in \bi}\beta_{}^{i_2'}\alpha_{i_1'i_2'}\right )X_{ni_1'}}{\sqrt{\mbox{Var}(\tilde e_{n}^{\bi})}}\right )\\
&=\frac{\exp\{\widetilde{\beta}^{\bi\top}X_{n}^{\bi}\}}{1+\exp\{\widetilde{\beta}^{\bi\top}X_{n}^{\bi}\}},
\label{eq:rescaleproof}
\end{align*}
 where $\widetilde{\beta}^{\bi} = \frac{ {\beta}^{\bi}+ 
\left[ 0, \left( \sum_{i' \in \mathcal{I}_1^{-\bi}} \beta^{i'} \alpha_{\bi i'}\right)^\top \right]^\top }{\sqrt{\mbox{Var}(\tilde e_{n}^{\bi})}}\in \real^{(j+1)}$. Therefore we see that the regression parameters $\widetilde{\beta}^{\bi}$ for the OvR model fit to the data indexed by $\bi $ can be expressed as functions of the regression parameters $\beta$ for the complete $D-$dimensional OvR model. The value for $\widetilde{\beta}^{\bi}$ that maximises the binary logistic likelihood for the $\bi^{th}$ subset is therefore a rescaled version of the value for $\beta^{\bi}$ that maximises the complete binary logistic likelihood. 

\cite{Adelin1984} proved that if the dataset $\bx$ does not exhibit complete or quasi-complete separation, then there is a unique value for the MLE $\hat\beta$ for the complete model regression parameter $\beta$. As a result, unique values for the MLE $\hat{\beta}^{\bi}$ for $\widetilde{\beta}^{\bi} $ exist if the data does not exhibit complete separation in the variables indexed by $\bi$. It is trivial to show that if the complete dataset $\bx$ does not exhibit complete or quasi-complete separation, then there is no $\bi\in I_j$ for which $\bx^{\bi}$ exhibits complete or quasi-complete separation. If there was, then there would exist a $b^{\bi}$ such that 
\begin{align*}
b^{\bi\top} x_n^{\bi}& > 0 \mbox{ for all } n \mbox{ such that } y_n=2\\
b^{\bi\top} x_n^{\bi}& < 0 \mbox{ for all } n \mbox{ such that } y_n=1.
\end{align*}
Setting $b_{d}=b_{d}^{\bi}$ if $d\in\bi$ and $0$ otherwise yields the vector $b=(b_{1},\ldots,b_{D})$ such that
\begin{align*}
b^{\top} x_n=b^{\bi\top} x_n^{\bi}& > 0 \mbox{ for all } n \mbox{ such that } y_n=2\\
b^{\top} x_n=b^{\bi\top} x_n^{\bi}& < 0 \mbox{ for all } n \mbox{ such that } y_n=1,
\end{align*}
which is a contradiction. As a result, a sufficient condition for the existence and uniqueness of all MLE's $\hat{\beta}^{\bi}$ for $\widetilde{\beta}^{\bi}$, $\bi\in\bi$, is that the complete data $\bx$ does not exhibit complete or quasi-complete separation. Now, given $\hat{\beta}^{\bi}$ is the value for $\widetilde{\beta}^{\bi}$ that minimises the log-likelihood for the $\bi^{th}$ model
$$\log L(\bx^{\bi},y;\widetilde{\beta}^{\bi})=\sum_{n=1}^N\boldsymbol{1} \{y_n=1\}\log P(Y=1|x_n^{\bi},y_n,\widetilde{\beta}^{\bi})+\boldsymbol{1} \{y_n=2\}\log P(Y=2|x_n^{\bi},y_n,\widetilde{\beta}^{\bi}),$$
i.e.~$\log L(\bx^{\bi},\by;\hat{\beta}^{\bi})<\log L(\bx^{\bi},\by;\widetilde{\beta}^{\bi})$ for all $\widetilde{\beta}^{\bi} \in \dom_{\widetilde{\beta}^{\bi}}$, the parameter $\hat \bbeta^{(j)}=\left \{\hat{\beta}^{\bi}\right \}_{\bi\in I_j}$ therefore minimises the log-likelihood
$$\log L_{\tO}^{(j)}(\bx, \by; \widetilde\bbeta^{(j)}) = \sum_{\bi \in I_j}\log L(\bx^{\bi},\by;\tilde\bbeta^{\bi}),$$
i.e.~$\log L_{\tO}^{(j)}(\bx, \by; \hat\bbeta^{(j)})<\log L_{\tO}^{(j)}(\bx, \by; \widetilde\bbeta^{(j)})$ for all $\widetilde{\bbeta^{(j)}}^{} \in \dom_{\widetilde{\bbeta^{(j)}}^{}}$. Therefore an estimate $\hat\beta$ for $\beta$ such that
$$\hat{\beta}^{\bi} = \frac{ {\beta}^{\bi} + 
\left[ 0, \left( \sum_{i' \in \mathcal{I}_1^{-\bi}}\hat \beta^{i'} \alpha_{\bi i'}\right)^\top \right]^\top }
{\sqrt{1 + \frac{\sum_{i_1' \in \mathcal{I}_1^{-\bi}} \left( \beta^{i_1'2} \lambda_{\bi i_1'}^2 + 2  \sum_{i_2' \in \mathcal{I}_1^{-\bi}, i_2' \neq i_1'} \beta^{i_1'} \beta^{i_2'} \lambda_{\bi i_1' i_2'} \right)}{\pi^2/3}}},$$
(i.e.~an estimate for $\beta$ that yields the MLE's for each of the smaller-dimensional logistic models based on the variables indexed by $\bi$) will minimise $\log L_{\tO}^{(j)}(\bx, \by; \widetilde\bbeta^{(j)})$. By the definition of the model this $\hat\beta$ will exist. Given that a logistic OvR model is just the product of $K$ binary logistic regression models, the above results hold for the OvR model for $K>2$ classes. Consequently, the $j-$dimensional OvR model can therefore be written as 
$$L_{\tO}^{(j)}(\bx, \by; \bbeta) = \prod_{\bi \in \mathcal{I}_j} L_{\tO} (\bx^{\bi}, \by; \widetilde{\bbeta}^{\bi}),$$
where 
$$\widetilde{\beta}^{\bi}_k = \frac{ {\beta}^{\bi}_k + 
\left[ 0, \left( \sum_{i' \in \mathcal{I}_1^{-\bi}} \beta_k^{i'} \alpha_{\bi i'}\right)^\top \right]^\top }
{\sqrt{\mbox{Var}(\tilde e_{nk}^{\bi})}}\in \real^{(j+1)}.$$

Through similar arguments, we can find an expression for the symbolic $j$-dimensional OvR model as
\begin{equation*}
L_{\tSO}^{(j)}(\bs; \bbeta) = \prod_{\bi \in \mathcal{I}_j} L_{\tSO} (\bs^{\bi}, \by; \widetilde{\bbeta}^{\bi}).
\end{equation*}

\bibliographystyle{chicago}
\bibliography{bib}

\end{document}